\documentclass[twocolumn]{aastex63}
\usepackage{color}
\usepackage{listings}
\usepackage{url}
\usepackage{verbatim}
\usepackage{subfigure}
\usepackage{amsmath}
\usepackage{enumitem}
\usepackage{float}
\usepackage{graphicx}

\setlist[enumerate]{label*=\arabic*.}
\graphicspath{{figures/}}

\newcommand{\code}[1]{\texttt{#1}}

\definecolor{dkgreen}{rgb}{0,0.6,0}
\definecolor{gray}{rgb}{0.5,0.5,0.5}
\definecolor{mauve}{rgb}{0.58,0,0.82}
\shorttitle{Photometric Classification of SN~1991bg-like Supernovae in SDSS-II}
\shortauthors{Perrefort et al.}

\begin{document}

\title{A Template-Based Approach to the Photometric Classification of SN~1991bg-like Supernovae in the SDSS-II Supernova Survey}

\correspondingauthor{Daniel Perrefort}
\email{djperrefort@pitt.edu}

\author[0000-0002-3988-4881]{Daniel Perrefort}
\affiliation{
    Pittsburgh Particle Physics, Astrophysics, and Cosmology Center (PITT PACC).
    Department of Physics and Astronomy,
    University of Pittsburgh,
    Pittsburgh, PA 15260, USA
    }
    
\author{Yike Zhang}
\affiliation{
    Pittsburgh Particle Physics, Astrophysics, and Cosmology Center (PITT PACC).
    Department of Physics and Astronomy,
    University of Pittsburgh,
    Pittsburgh, PA 15260, USA
    }

\author[0000-0002-1296-6887]{Llu\'is Galbany}
\affiliation{
    Pittsburgh Particle Physics, Astrophysics, and Cosmology Center (PITT PACC).
    Department of Physics and Astronomy,
    University of Pittsburgh,
    Pittsburgh, PA 15260, USA
    }
 \affiliation{
Departamento de F\'isica Te\'orica y del Cosmos, Universidad de Granada, E-18071 Granada, Spain.
    }

\author[0000-0001-7113-1233]{W.~M. Wood-Vasey}
\affiliation{
    Pittsburgh Particle Physics, Astrophysics, and Cosmology Center (PITT PACC).
    Department of Physics and Astronomy,
    University of Pittsburgh,
    Pittsburgh, PA 15260, USA
 }
 
 \author[0000-0001-9541-0317]{Santiago Gonz\'alez-Gait\'an}
\affiliation{
CENTRA-Centro de Astrof\'isica e Gravitaç\~ao and Departamento de F\'isica, Instituto Superior T\'ecnico, Universidade de Lisboa, Avenida Rovisco Pais, 1049-001 Lisboa, Portugal.
}

\received{Jan 17, 2020}
\revised{Sep 4, 2020}
\accepted{Sep 12, 2020}
\submitjournal{The Astrophysical Journal}
\reportnum{Manuscript: AAS 22419}

\begin{abstract}
The use of Type Ia Supernovae (SNe~Ia) to measure cosmological parameters has grown significantly over the past two decades. However, there exists a significant diversity in the SN~Ia population that is not well understood. Over-luminous SN~1991T-like and sub-luminous SN~1991bg-like objects are two characteristic examples of peculiar SNe. The identification and classification of such objects is an important step in studying what makes them unique from the remaining SN population. With the upcoming Vera C. Rubin Observatory promising on the order of a million new SNe over a ten-year survey, spectroscopic classifications will be possible for only a small subset of observed targets. As such, photometric classification has become an increasingly important concern in preparing for the next generation of astronomical surveys. Using observations from the Sloan Digital Sky Survey II (SDSS-II) SN Survey, we apply here an empirically based classification technique targeted at the identification of SN~1991bg-like SNe in photometric data sets. By performing dedicated fits to photometric data in the rest-frame redder and bluer bandpasses, we classify 16 previously unidentified 91bg-like SNe. Using SDSS-II host-galaxy measurements, we find that these SNe are preferentially found in host galaxies having an older average stellar age than the hosts of normal SNe~Ia.  We also find that these SNe are found at a further physical distance from the center of their host galaxies.
We find no statistically significant bias in host galaxy mass or specific star formation rate for these targets.
\end{abstract}
\section{Introduction} \label{sec:introduction}

Following their fundamental role in the discovery of the accelerating expansion of the Universe \citep{Riess98, Perlmutter99}, Type Ia Supernovae (SNe~Ia) have been used to determine cosmological parameters with an increasing level of accuracy and precision \citep{Betoule14, Rest14, Scolnic14, Scolnic17, Jones18, DES18}. The use of SNe~Ia as cosmological probes relies on the fact that SN~Ia luminosities at the time of maximum brightness are not only bright but also have low intrinsic scatter. This scatter can be reduced even further by calibrating their intrinsic peak luminosity with light-curve width \citep{Phillips93, Phillips99} and optical color \citep{Riess96, Tripp99}. However, even after applying these corrections SNe~Ia remain a heterogeneous collection of objects spanning a diverse collection of subtypes \citep{Taubenberger17}. 

Early attempts at classifying peculiar SNe~Ia quickly identified categories of overluminous, SN~1991T-like objects \citep{Filippenko92_91t, Phillips92} and subluminous, fast-declining objects like SN~1991bg \citep{Filippenko92_91bg, Leibundgut93, Turatto96}. More recent works have introduced additional classifications based on spectroscopic properties such as SN~2002es-like SNe \citep{Ganeshalingam12}, which are sub-luminous but slow declining, super-Chandrasekhar-mass candidates \citep{Howell06}, 2002cx-like SNe, also known as SNe~Iax \citep{Li03, Jha06, Meng18}, and others with fewer observed targets. However, the presence of SNe like SN~1991T and SN~1991bg make up the predominant population of observed peculiar SNe.

There is some disagreement in the literature when it comes to the rate of 91bg-like events. Recent rate estimates using SNe observed by the Lick Observatory Supernova Search (LOSS) range from 11 to 15\% of the SN~Ia population \citep{Ganeshalingam10, Li11}. Alternatively, \cite{Gonzalez_Gaitan11} and \cite{Silverman12} estimate that 91bg-like SNe make up a more modest 6 to 9\% using various, low-redshift data sets. However, \cite{Gonzalez_Gaitan11} notes that their estimates increase dramatically with the inclusion of transitional SN~1986G-like SNe, which have luminosities that lie in the intermediate range between normal and SN~1991bg-like.

The upcoming Vera C. Rubin Observatory will conduct the Legacy Survey of Space and Time\footnote{Previously known as the Large Synoptic Survey Telescope. \\ \url{https://www.lsst.org/news/vro-press-release}} \citep[LSST;][]{LSST09} and observe hundreds of thousands of new SNe over a ten-year survey, promising a dramatic increase in the number of observed peculiar SNe. However, the availability of spectroscopic follow-up observations and, as a result, spectroscopically determined classifications, will be heavily limited. The ability to provide accurate, photometric classifications will thus be increasingly important in the coming years for maximizing the science that can be done with the Rubin Observatory. 

One approach to this challenge is the development of machine learning classifiers designed to reproduce existing classification schemes \citep[e.g.,][]{Richards12, Karpenka13, Varughese15, Lochner16, Moller16, Sasdelli16, Dai18, Muthukrishna19, Pasquet19}. Although machine learning classifiers benefit from the ability to scale to large data sets, they don't reveal the underlying physics that lead to a classification. The ability of a machine learning classifier to identify unexpected, peculiar objects is also extremely sensitive to the quality and diversity of the initial training sample.

An alternative is to classify SNe based on their light-curve properties. Empirically based classification schemes are not only transparent in how they work, but can simultaneously provide physically motivated values such as light-curve color, standardized peak luminosities, and decline rates. Furthermore, many SN analyses already employ the use of a light-curve fitter, making it easy to incorporate classification into existing analysis procedures. 

In \citet[][G14 hereafter]{Gonzalez-Gaitan14} a photometric identification technique was introduced for discriminating SN~1991bg-like objects in photometric samples. Using several low-redshift samples from the literature, G14 demonstrated that this method is not only capable of identifying dim, fast-declining SNe, but can potentially identify other peculiar transients such as SNe~Iax-like, SN~2006bt-like, and super-Chandrasekhar SNe~Ia. 
We apply here the same classification technique to a larger target sample and compare results against spectroscopically determined subtypes. 

We make two significant changes to the original approach of G14. The first is the use of a newer model for 91bg-like SNe that has been extended further into the near infra-red (NIR) and ultraviolet (UV). By using this extended model, we are able to apply the classification to a larger, higher-redshift sample of SNe~Ia. Secondly, we consider multiple implementations of the technique and discuss potential biases that may arise.

The layout of this paper is as follows: In Section \ref{sec:data} we discuss the supernova sample considered by this paper. In Section \ref{sec:classification_method} we present our chosen classification method, including a detailed outline of our procedure in Section \ref{ssec:classification_procedure} and the models employed by our analysis in Section \ref{ssec:sne_models}. 
Our photometric classifications are then presented in Section \ref{sec:photometric_classification} followed by our conclusions in Section \ref{sec:conclusion}. A complete technical outline of this work, including Python code suitable for reproducing results, is available online.\footnote{See \url{https://Perrefort2020.readthedocs.io/en/latest/}}
\section{Data} \label{sec:data}

In this work, we focus on the photometric classification of SNe observed by the Sloan Digital Sky Survey II (SDSS II) Supernova Survey. Since spectroscopic followup is limited for the SDSS SN sample, we additionally consider spectrophotometric observations taken by the Carnegie Supernova Project I (CSP-I), allowing us to better estimate the performance of our classification technique.

\subsection{SDSS-II}

\begin{figure}[t]
  \includegraphics[width=\columnwidth]{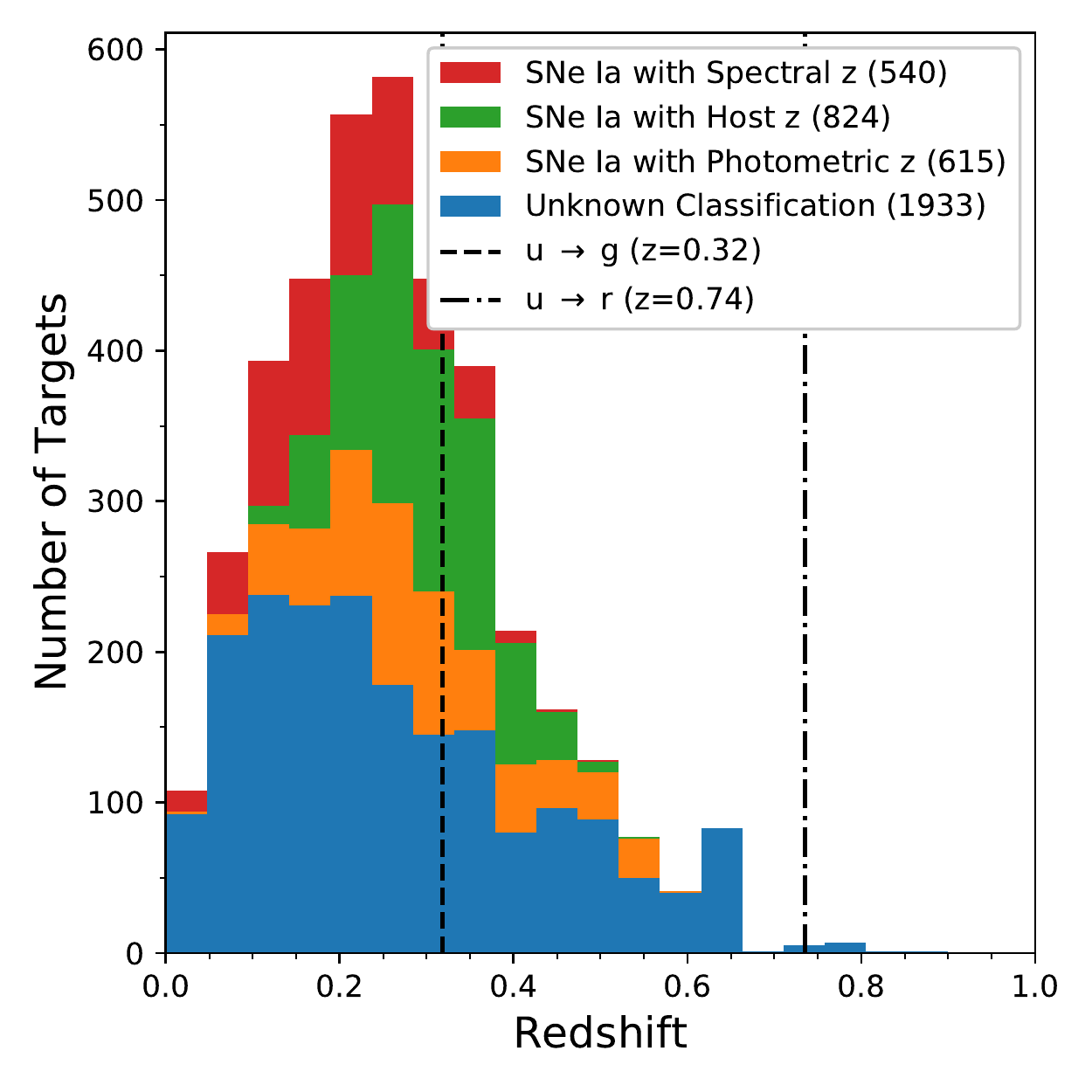}
  \caption{Stacked redshift distribution for the objects considered in this paper, grouped by their classification in S18.
  Spectroscopic classifications are available predominately for lower redshift targets, while higher redshift targets have fewer photometric or spectroscopic classifications.
  Reference lines mark redshift values at which rest-frame \textit{u}-band observations are equivalent to the observer frame \textit{g} and \textit{r} band as determine by effective wavelength. Not included are 76 objects with unknown classifications for which a redshift value could not be determined.}
  \label{fig:sdss_redshift}
\end{figure}

The SDSS II SN survey \citep[][S18 hereafter]{Sako18} was performed using the 2.5-m Sloan Foundation Telescope \citep{York00, Gunn06} at Apache Point Observatory (APO). The SDSS II SN survey ran over three observing seasons from 2005 to 2007, covering a 300 square-degree stripe of sky along the celestial equator in the Southern Galactic hemisphere in the \textit{ugriz} bands \citep{Fukugita96, Doi10}.
All SDSS SNe are refered to in this work using their associated Candidate Identifier (CID) published in S18.

The SDSS SN data release provides light-curve data for 10,258 variable and transient sources. This includes 540 objects spectroscopically classified as SN~Ia, 64 SN II, 3 super-luminous SN (SLSN), 22 objects classified as either SN Ib or Ic, and 4,131 sources that are either variable or AGNs. Additionally, there are 2,009 targets with light curves that were deemed too sparse or noisy to provide a classification. Shown in Figure \ref{fig:sdss_redshift}, this combined sample spans redshifts out to $z \leq 0.9$. 

Initial classification of the SDSS spectra were performed in \cite{Zheng08} using the \code{rvsao.xcsao} cross-correlation package of IRAF \citep{Tody93}.
Additional photometric classifications using the SDSS photometry were performed by \cite{Sako11} using an extension of the Photometric SN IDentification (PSNID) software \citep{Sako08}. 
Further delimitations to each classification were added manually by \cite{Sako11} depending on whether the classification was made photometrically (denoted with a prefix \textit{p}) spectroscopically (no prefix) or using a host galaxy redshift (prefix \textit{z}). These classifications are referred to throughout this work as a reference for those familiar with the SDSS data set or other uses of PSNID. 

No systematic search or sub-typing effort for peculiar SNe was performed beyond the assignment of basic SN types \citep{SakoPrivComm}. However, a small selection of targets was manually flagged based on their spectroscopic or photometric properties. These SNe are listed in Table \ref{tab:sdss_peculiars} and include 4 SNe possibly similar to SN~1991bg, 1 SN possibly similar to SN~2000cx, 2 SNe possibly similar to SN~2002ci, and 3 SNe possibly similar to SN~2002cx. 

Photometric zero points for SDSS-II were determined using stars from the Ivezi\'c catalog \citep{Ivezic07}. The position, band-specific flux, and host-galaxy intensity for each target were then fitted for using Scene Model Photometry \citep[SMP;][]{Holtzman08}. 
\citet{Betoule13} provides the most updated calibration of these data using a position-dependent correction, which were applied in S18 and are also used in this work.

\subsection{CSP-I}

CSP-I was a five-year survey in the optical and NIR run at Las Campanas Observatory (LCO) from 2004 through 2009. 
Optical observations were taken in the \textit{ugriBV} bandpasses \citep{Stritzinger11} using the SITe3 and Tek5 CCD cameras on the Swope 1 m and du Pont 2.5 m telescopes respectively. NIR imaging was performed in the \textit{YJH} bands \citep{Contreras10} using the Wide-Field IR Camera (WIRC) \citep{Persson02} on the du Pont 2.5 m telescope and later RetroCam on the Swope 1 m telescope.

CSP-I includes spectrophotometric observations of 134 SNe Ia spanning $z < 0.085$ \citep{Hamuy06, Krisciunas17}. Out of these targets, 96 were classified as normal SNe Ia, 13 as being like SN 1991bg, 5 like SN 1991T, 5 like SN 2002cx, and 4 were unclassified. 

In \cite{Mosher12} overlapping observations between CSP-I and SDSS-II were compared for 9 spectroscopically confirmed SNe Ia, five of which were classified as peculiar. Photometric observations in the \textit{gri} bands were found to agree within a 1\% level in flux with a typical epoch-to-epoch scatter no greater than 0.05 magnitudes. The \textit{u} band scatter was slightly higher at 0.077 magnitudes, but flux values were still consistent within 1\%. Taking into account the small sample size along with systematic uncertainties in the analysis, it was estimated that offsets in observer frame \textit{u} were conservatively within 0.04 magnitudes.

\section{Classification Method} \label{sec:classification_method}

\begin{figure*}
 \centering
 \includegraphics[width=.9\textwidth]{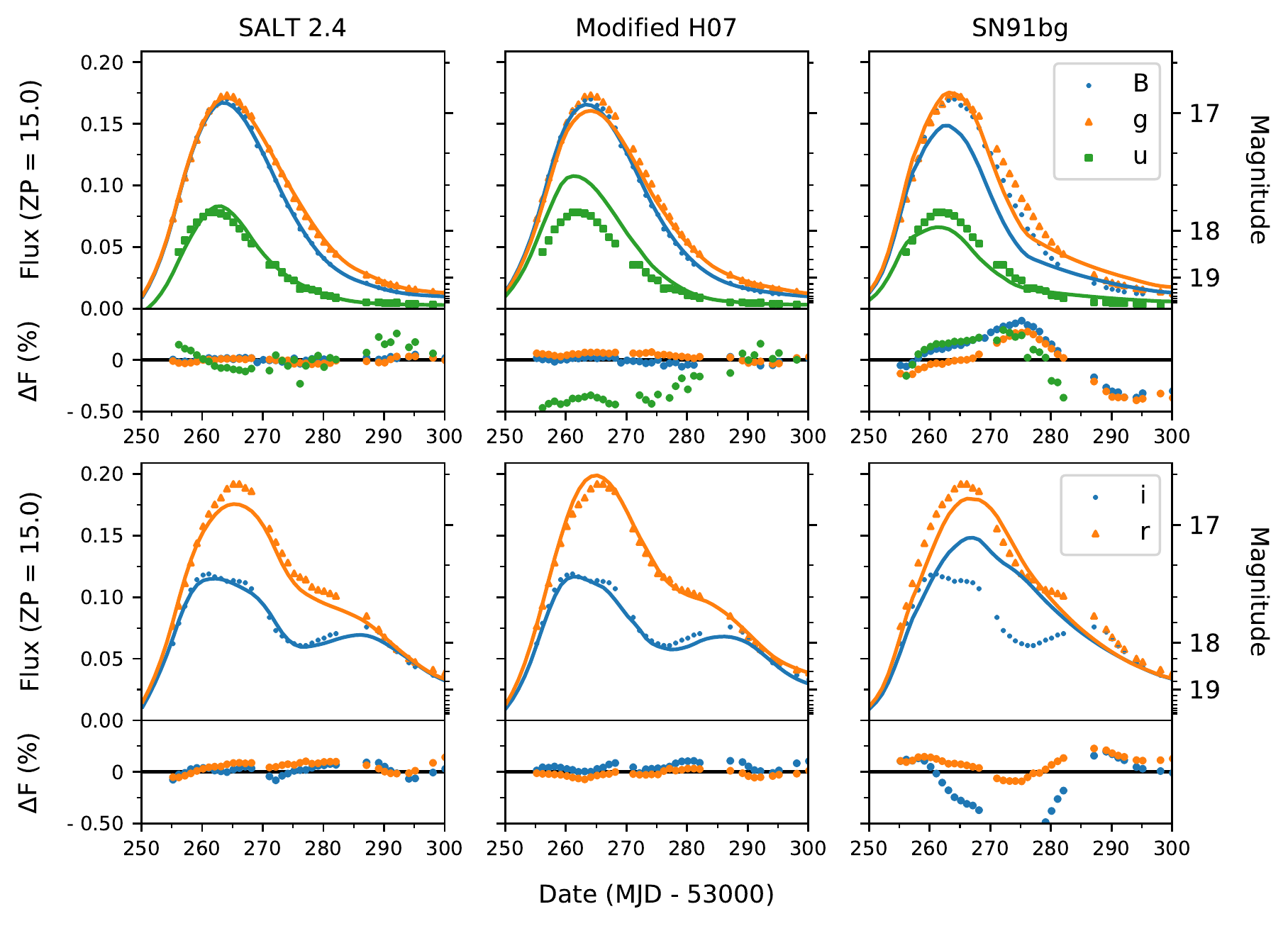}
 \caption{CSP observations of the spectroscopically normal SN~2004ef are fit separately in the rest-frame blue (top) and red (bottom) bands using the SALT 2.4 (left), H07 (middle), and SN~1991bg-like (right) models. We note in the blue bands that the fitted 91bg model is narrower and fainter at peak than the observations. We also note that the H07 model, which is trained using a more heterogeneous set of template spectra than SALT 2.4, overestimates the $u$-band. In the red bands we see that the morphology of each model plays a greater role. In particular, the lack of a secondary maximum in the 91bg-like model lends greater importance to late time observations (phase $\gtrsim 30$ days) in constraining the width of the modeled light-curve.
}
 \label{fig:fit_of_normal_sn}
\end{figure*}

\begin{figure*}
 \centering
 \includegraphics[width=.9\textwidth]{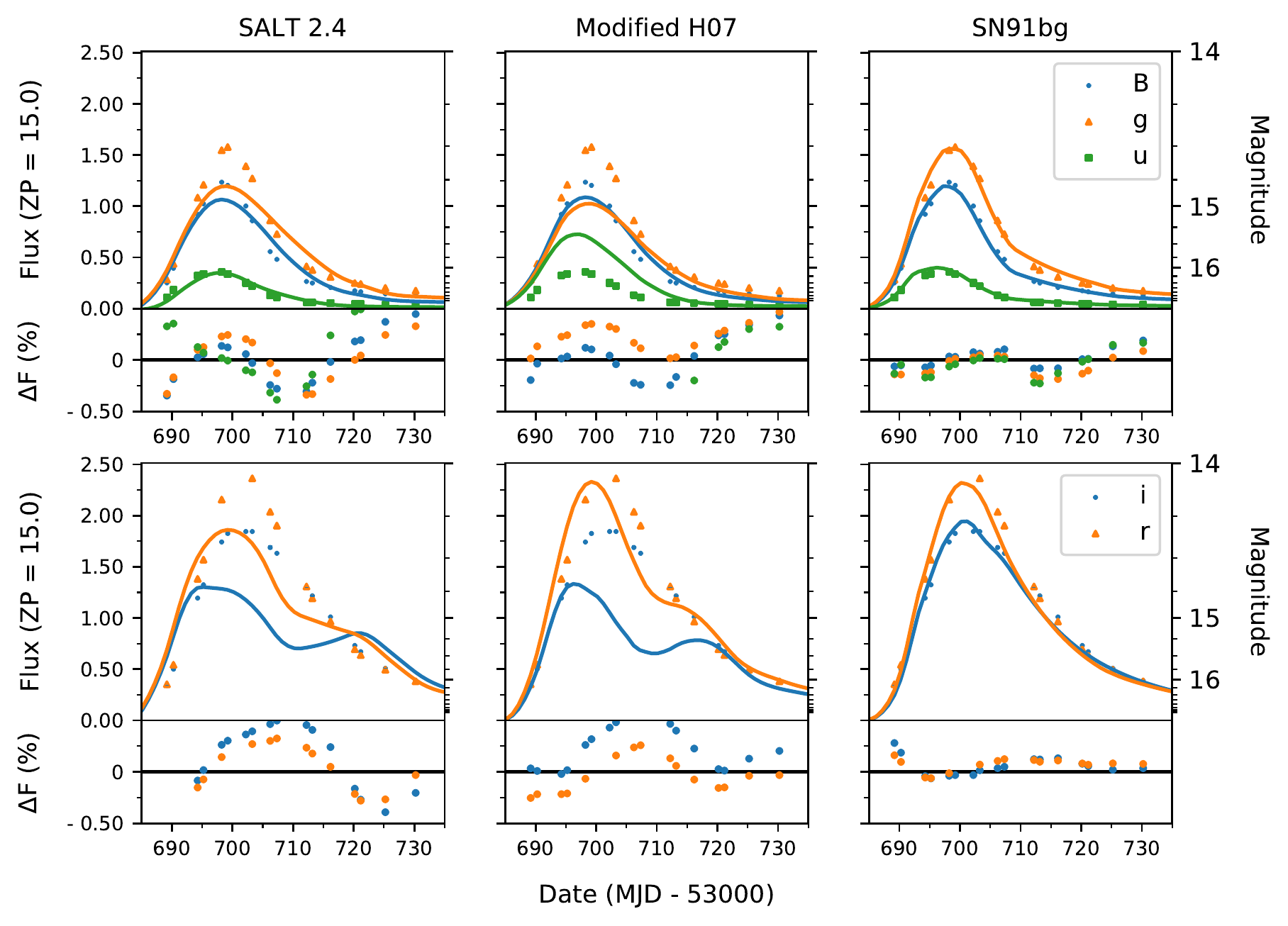}
 \caption{CSP observations of the 91bg-like SN~2005ke are fit separately in the rest-frame blue (top) and red (bottom) bands using the SALT 2.4 (left), H07 (middle), and 91bg-like (right) models. We note that the SALT 2.4 and H07 fits --- which represent normal SNe~Ia --- are significantly bluer when fit in the \text{ugB} bands than the SN~1991bg model which is a significantly better fit to the data.}
 \label{fig:fit_of_91bg}
\end{figure*}

The classification of 91bg-like SNe relies on them having distinct spectrophotometric differences from the normal SN~Ia population. Most notably, the presence of strong Ti lines in their spectra indicates that 91bg-like SNe tend to be cooler than normal SNe~Ia, which can generally be attributed to a lower yield of $^{56}\text{Ni}$ synthesized in the explosion \citep{Nugent95}. In photometric terms, this means 91bg-like events have light-curves that are both fainter at maximum than normal SNe~Ia (M$_\text{B} (\text{91bg}) \sim -18$) and decline more rapidly after peak brightness. The latter of these effects is typically well demonstrated by the light-curve parameters stretch or $\Delta m_{15}$, where typical values are found to be $\Delta m_{15}(B) \gtrsim 1.7$ \citep{Galbany19}.

The cooler explosion temperature of 91bg-like SNe also has an impact on the evolution of light-curve color. The lower temperatures allow for the recombination of Fe III to Fe II to happen sooner, which results in redder colors at maximum. This causes the epoch of peak brightness for redder bands to be delayed when compared with the normal SNe~Ia population.  SN~1991bg-like explosions also lack the secondary maximum seen in the redder bands in normal SNe~Ia.  The delay in the peak brightnesses combined with the lack of a shoulder or secondary maxima in redder bands make 91bg-like SNe identifiable by their photometric properties.

% This more general classification method will allow us to identify more SNe 91bg, but has the potential to misclassify more normal SNeIa with intrinsically lower stretch or redder colors. For example, highly reddened, normal SNe with a low stretch could look potentially similar 91bg-like SNe in terms of $\chi^2$.

In principle, one might attempt to distinguish 91bg-like events by directly looking for these photometric properties --- specifically the lack of a secondary maximum. However, this requires an observational cadence with even and complete sampling from maximum light through the secondary maximum.
These problems can be alleviated by performing an overall fit to  the  data  and  selecting  targets  based  on  a $\chi^2$ value or a set of model parameters, but this raises additional challenges since there may be some normal SNe with intrinsically lower stretch or redder colors.  For example, highly reddened, normal SNe with a low stretch would look potentially similar to 91bg-like SNe in terms of $\chi^2$. 
We here instead analyze SDSS-II SN data using the classification method of G14.

\subsection{Classification Procedure} \label{ssec:classification_procedure}

We employ our classification method following an adaptation of G14. To start, photometric observations for each target are split into collections of rest-frame red and blue bandpasses as defined by the rest-frame effective wavelength of each band $\lambda_{\rm z, eff}$. When doing so, we define \textit{blue} bandpasses as having $\lambda_{\rm z, eff} < 5,500$ \AA\, and \textit{red} bandpasses as having $\lambda_{\rm z, eff} > 5,500$ \AA. The use of 5,500 \AA\ is chosen to separate the rest-frame $ug$ bands from the rest-frame $riz$ and thus puts the secondary maximum in the \text{red} bandpasses.

We separately fit each of the blue and red data using two light-curve models: one representing normal SNe~Ia and one representing 91bg-like objects.
We calculate the $\chi^2$ for each combination based on the modeled flux $F$ for a set of parameters $\bar p$, the observed flux $f$, and the of degrees of freedom, $d=N-{\rm len}(\bar p)$.
\begin{equation}
\chi^2(\bar p) = \frac{1}{d}\sum_{i=1}^{N} \frac{(F(\bar p)-f_{i})^2}{\sigma_{i}^2} \, .
\end{equation}
Using the resulting $\chi^2$ values, we classify targets based on their position in the following phase space:
\begin{align} \label{eq:classification_coords}
    x &\equiv \chi^2_{\rm blue}(\text{Ia}) - \chi^2_{ \rm blue}(\text{91bg}) \, ,\\
    y &\equiv \chi^2_{\rm red}(\text{Ia}) \  - \chi^2_{ \rm red}(\text{91bg}).
\end{align}
By construction of the above coordinates, we expect 91bg-like SNe to fall in the upper right (first) quadrant while normal SNe~Ia should fall in the lower left (third) quadrant of the $x, y$ plane.

Our normal SNe~Ia and 91bg-like models are imperfect, particularly in the model variances.  Thus the classification quadrants may not be best separated by intersecting lines at $(0,0)$. An alternative origin may instead provide a higher level of purity when classifying peculiar SNe~Ia. Following G14, the quadrant boundaries are determined by using spectroscopically classified targets to maximize the figure of merit (FOM) parameter
\begin{equation} \label{eq:fom}
 {\rm FOM} = \frac{N_{\rm true}}{N_{\rm total}} \frac{N_{\rm true}}{N_{\rm true} + N_{\rm false}} \, ,
\end{equation}
where $N_{\rm total}$ represents the total number of objects with a given type (e.g., 91bg-like objects), $N_{\rm true}$ is the number of objects correctly classified as a given type, and $N_{\rm false}$ is the number of objects falsely classified as a given type. 

\subsection{Fitting Procedure} \label{ssec:fitting_procedure}

We choose to use the SNCosmo Python package to handle light-curve fits since it allows us to easily implement, modify, and apply a variety of template-based models \citep{Barbary16}. Unless otherwise stated, we use the \code{iminuit} minimization routine (\code{sncosmo.fit\_lc}) to determine best-fit parameters. By default, the SNCosmo package fits each model using a single set of global, model-dependent parameters. However, we note that this behavior is significantly different from the original implementation of the classification technique in G14.

In G14, SNe were classified using the SiFTO light-curve fitter, which is an empirical fitter that uses magnitude, light curve shape (i.e., stretch), and color to fit a light-curve \citep{Conley08}. It is important to note that SiFTO uses band-specific flux normalizations as opposed to a set of global, light-curve specific parameters. Instead of relying on a dedicated color term, SiFTO allows the scale factor of the template to vary independently for each band. Similarly, the shape of the SiFTO model is described by a single stretch parameter that is applied differently in each observed filter as a function of effective wavelength. 

The choice of how parameters are varied across bands can potentially have a major impact on the resulting classification. In principle, we expect fits performed to red and blue data as collective sets to be more constrained by intrinsic color. However, by allowing parameters to vary across bands, the overall morphology of the light-curve is allowed to have a more significant impact. The drawback to this approach is a potentially higher sensitivity to the cadence of the observations. To understand the impact of this choice, we implement fitting routines for both approaches and compare the results. 

Our resulting fitting procedure is as follows:

\begin{enumerate}
 \item For each target, the Milky Way extinction is determined using the \cite{Schlegel98} dust map and the \cite{Fitzpatrick99} extinction law. This value is never varied in any fit.
 
 \item To determine a fiducial set of fit parameters, each light-curve is fit using both the H07 and SN91bg models and all available data points. At this step, all model parameters are varied except the redshift, which is only varied if it has not been determined with a spectroscopic observation.
 
 \item Using the redshift value determined in the previous step, the observed bandpasses are separated into the rest-frame blue ($\lambda_{\rm z, eff} < 5,500$ \AA) and red ($\lambda_{\rm z, eff} > 5,500$ \AA) bandpasses. 
 
 \item The red and blue data sets are fit using both models. So that we can investigate the resulting effect, we perform fits twice: once allowing fit parameters to vary independently across bands, and again using a single set of parameters for each red / blue dataset. At this stage, the redshift and time of \textit{B}-band maximum remain fixed to the value determined in step 2.
 
\end{enumerate}

% \begin{equation*}
% \end{equation*}

% The way that SNcosmo fits light curves is to minimize the $\chi^2$ defined by:
% \chi^2(z,t_0,x_1,c,A)=\sum_{i=1}^{N_t} \sum_{j=1}^{N_f}
% \end{equation*}

\subsection{Supernova Models} \label{ssec:sne_models}

Although the classification scheme described above only requires two SN models, we consider three models in our analysis: two used for target classification and a third as a baseline reference for comparison with existing results in the literature. When discussing the general properties of an observed light-curve, we default to the fitted parameters of the SALT 2.4 model \citep{Guy07}. The remaining two models are chosen to closely mimic those used in G14.

For normal SNe, we use the spectral time series template from \citet[][H07 hereafter]{Hsiao07}. This template was intentionally constructed to incorporate a large and heterogeneous sample of observed spectra and is the same model used by G14. Although the H07 model is already built into SNCosmo, the default model does not include a stretch-like parameter. 
This is problematic for two reasons. Firstly, it lends the other models a potentially unfair advantage in their flexibility to fit a given light-curve. Secondly, it limits our ability to investigate the impacts of fitting each bandpass independently versus as red / blue sets. To address these issues we add a stretch parameter $-0.5 < x_1 < 0.5$ to the pre-existing parameters of amplitude $A$, redshift $z$, and time of \textit{B}-band maximum $t_0$. 

Here and throughout this paper, we have chosen the variable $x_1$ to represent a stretch-like parameter similar in significance to that of SALT 2.4. However, we note that the full meaning of this parameter is uniquely dependent on the model being discussed. For the H07 model, we implement the $x_1$ parameter such that the flux $F$ is determined from the template $f_\text{H}(t, \lambda)$ as:

\begin{equation} \label{eq:stretch_param}
 F_\text{H}(t, \lambda, z, A, t_0, x_1) = A f_\text{H}\left( \frac{t - t_0}{x_1 + 1}, \frac{\lambda}{(1+z)} \right) \, .
\end{equation}

As a model for 91bg-like SNe, we employ the same spectroscopic template used in the PLAsTiCC challenge \citep{Plasticc18}, which includes the same parameters as the H07 model in addition to a color parameter $c$. This template is based on the 91bg model from \cite{Nugent02} but is extended further into the NIR and ultraviolet (UV) using synthetic spectra from \cite{Hachinger08} and light-curves observed by Swift \citep{brown09}. This allows the model to cover a broader wavelength range from 1,000 to 12,000 \AA\, and permits the fitting of targets at higher redshift values. The parameters of the 91bg model span stretch values from $0.65 \leq x_1 \leq 1.25$ and color values from $0 \leq c \leq 1.0$.

Using observations from CSP \citep{Krisciunas17}, we compare fits of all three models to a spectroscopically normal and 91bg-like SN in Figures \ref{fig:fit_of_normal_sn} and \ref{fig:fit_of_91bg} respectively. We note in the blue bands that the most significant difference between the fitted models is their color, although the decline rate does play a secondary effect. Conversely, the biggest difference between the normal and 91bg-like models in the red bands is the overall morphology (i.e., the existence or lack of a secondary maximum). In particular, we see that late time observations ($\gtrsim$ 30 days past maximum) play an important role in constraining the stretch of the SN~1991bg model when fitting normal SNe~Ia light-curves. We also note that there is a key difference in the epoch of the first maximum, particularly in the blue bands, which plays an important role in determining the resulting chi-square.
\section{Photometric Classification Results} \label{sec:photometric_classification}

Table \ref{tab:class_fits} presents the fitted parameters for each SN using both the SN~1991bg and modified H07 models. Fits are performed for all targets not classified by S18 as being non-transients (i.e., as either Variable or AGN like objects). To ensure fits are well constrained by the data, we disregard any targets not having observations with a signal to noise ratio SNR $\geq 5$ in two or more bandpasses. Furthermore, we require at least one of these observations to fall between -15 and 0 days of the fitted SALT 2.4 \textit{B}-band maximum, and the other between 0 and 25 days. Targets not matching these criteria are dropped from our sample, leaving a total of 3,882 remaining targets.

\subsubsection{FOM Optimization}
\label{ssec:fom_optimization}

%   \centering
%   \caption{Distributions of the signal to noise ratio (SNR) for spectroscopically confirmed SNe~Ia observed by SDSS (left) and CSP (right). Seperate distributions are shown for observations taken in the rest-frame blue ($\lambda_{\rm z, eff} < 5,500$; solid blue) and the rest-frame red ($\lambda_{\rm z, eff} > 5,500$; outlined orange). Objects observed by CSP tend to have significantly higher SNR due to the lower redshift limit of the survey, with a median SNR that is $\sim$44 (33) times larger than SDSS in the rest-frame blue (red) bands.}
% \end{figure*}

In order to optimize Equation \ref{eq:fom}, we require a set of spectroscopically classified SNe~Ia. 
Spectroscopic classification of SDSS targets was attempted for this work following the prescription of \cite{Silverman12} using the SN IDentification software\footnote{Version 5.0: \url{https://people.lam.fr/blondin.stephane/software/snid/index.html}} \citep[SNID;][]{Blondin07}. Although excellent agreement was found with S18 when assigning SN types, subtyping results proved to be unreliable. Overall, this was attributed to either poor wavelength coverage, strong host galaxy contamination, and / or a low SNR in the observed spectra.

As an alternative, we suppliment our data set with spectroscopically classified objects from CSP \citep{Folatelli13}. By virtue of spanning a lower redshift range, photometric observations taken by CSP have, on average, a higher SNR than targets observed by SDSS. This is problematic since our classification scheme relies on a coordiante system that is based on chi-squared values and thus scales inversely with the average SNR (see Equation \ref{eq:classification_coords}). Our solution is to rescale the classification coordinates of CSP targets to more closly resemble SDSS using the median SNR in the rest-frame blue (SNR$_\text{B}$) and red bands (SNR$_\text{R}$) as follows:
\begin{align} \label{eq:scale_csp_coords}
    x'_{\text{CSP}} &\equiv \frac{{\text{SNR}}_{\text{B}, SDSS}}{{\text{SNR}}_{\text{B}, CSP}} \, x_{\text{CSP}},\\
    y'_{\text{CSP}} &\equiv \frac{{\text{SNR}}_{\text{R}, SDSS}}{{\text{SNR}}_{\text{R}, CSP}} \, y_{\text{CSP}}.
\end{align} 
Using the above definitions, we determine scale factors of 0.023 and 0.030 for the x and y coordinates respectively.

\begin{figure}
\centering  

\subfigure[Classification coordinates determined by fitting observed bandpasses independantly. Dashed lines are used to indicate classification boundaries $x=0.5$, $y=0$.]{\label{fig:band_fom_coords}\includegraphics[width=\columnwidth]{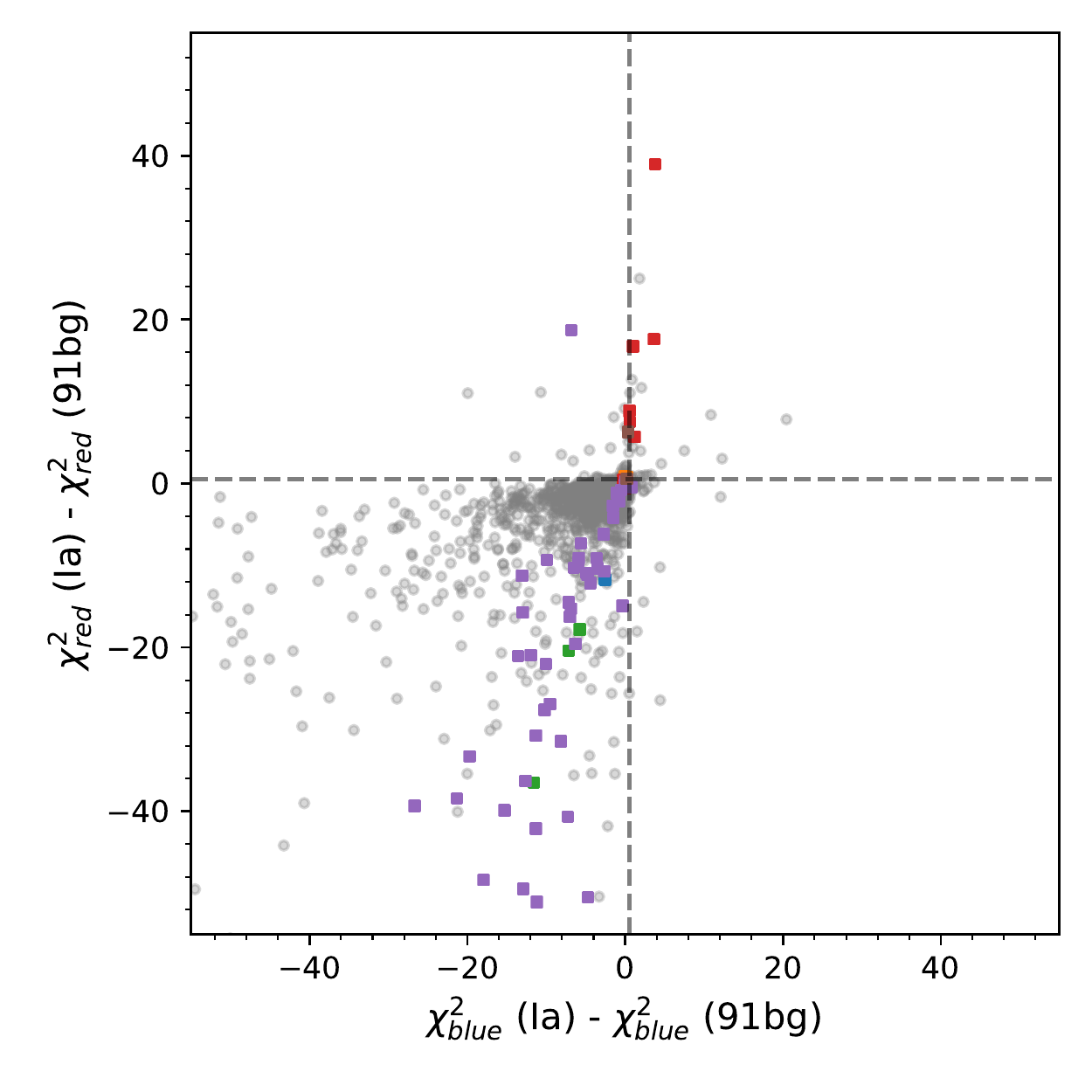}}

\subfigure[Classification coordinates determined by fitting rest-frame red / blue bandpasses as collective sets. Dashed lines are used to indicate classification boundaries of $x=3$, $y=0$.]{\label{fig:coll_fom_coords}\includegraphics[width=\columnwidth]{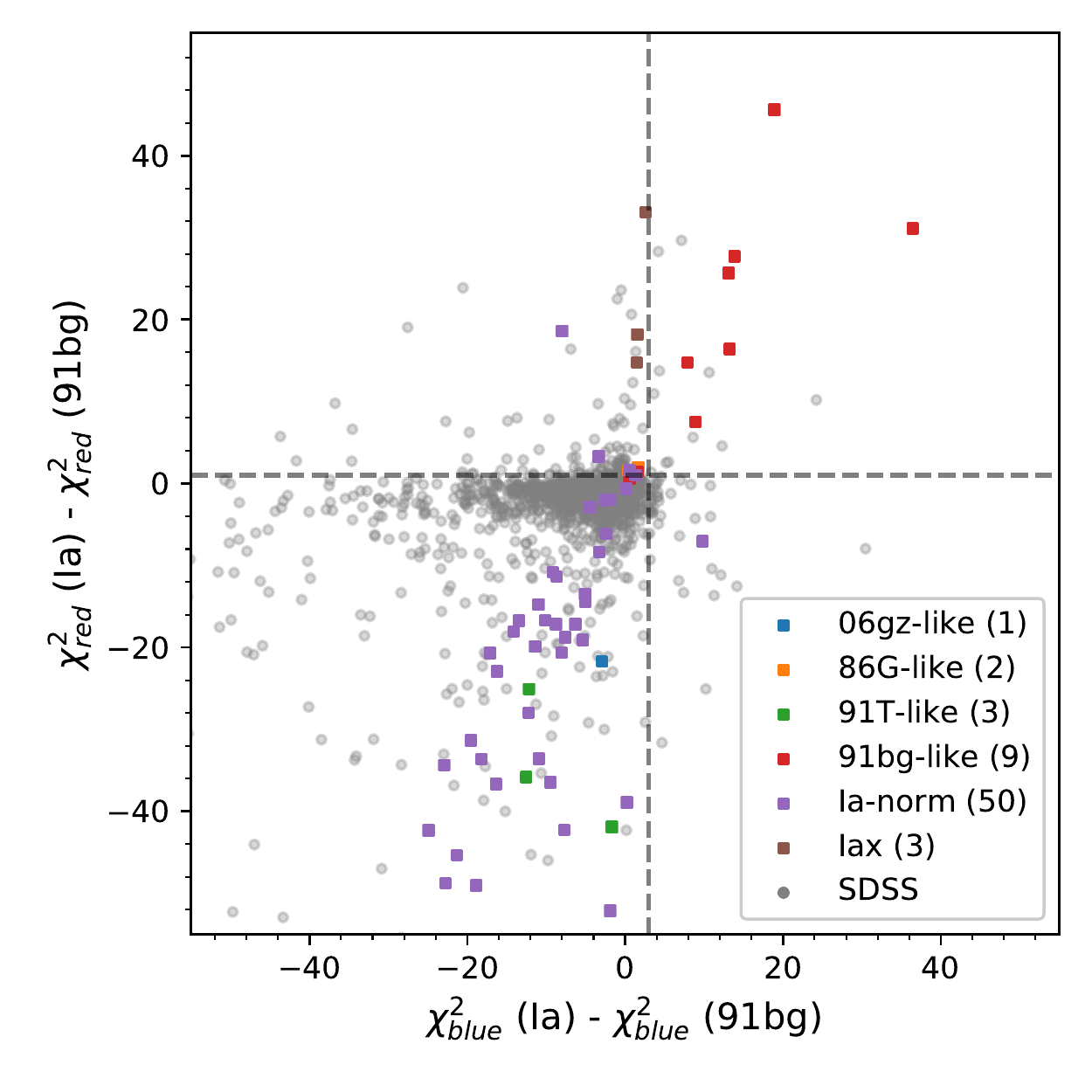}}

\caption{Classification coordinates are shown for objects observed by CSP (colored squares) and SDSS (grey circles) as determined by fitting rest-frame red / blue bandpasses independently (top) and as collective sets (bottom). CSP coordinates in both panels have been scaled by a factor of 0.023 along the x-axis and 0.030 along the y-axis to match the median SNR of SNe Ia observed by SDSS. The increased dispersion of points in the bottom panel indicates a lower sensitivity of the classification on the chosen classification boundaries.}
\label{fig:fom_coords}

\end{figure}

\begin{figure}
\centering  

\subfigure[FOM values determined by fitting observed bandpasses independantly. Dashed lines indicate classification boundaries of $x=0.5$, $y=0$ which align with the peak FOM value of $0.75$. FOM values are not available beyond $x \approx 4$ as no CSP SNe have classification cordinates in that region.]{\label{fig:band_fom_degeneracy}\includegraphics[width=\columnwidth]{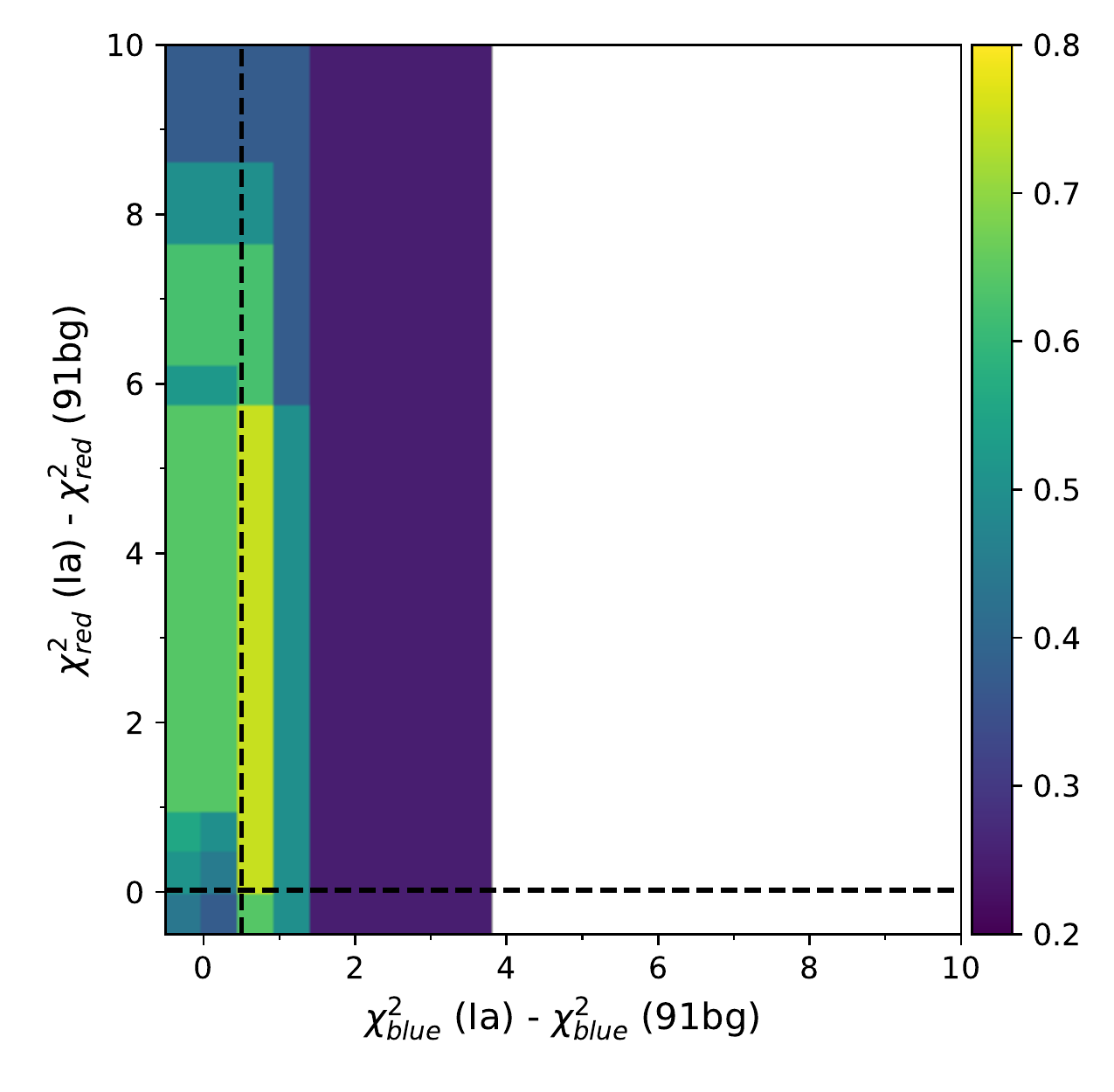}}

\subfigure[FOM values determined by fitting rest-frame red / blue bandpasses as collective sets. Dashed lines indicate classification boundaries of $x=3$, $y=0$ which align with the peak FOM value of $0.78$.]{\label{fig:coll_fom_degeneracy}\includegraphics[width=\columnwidth]{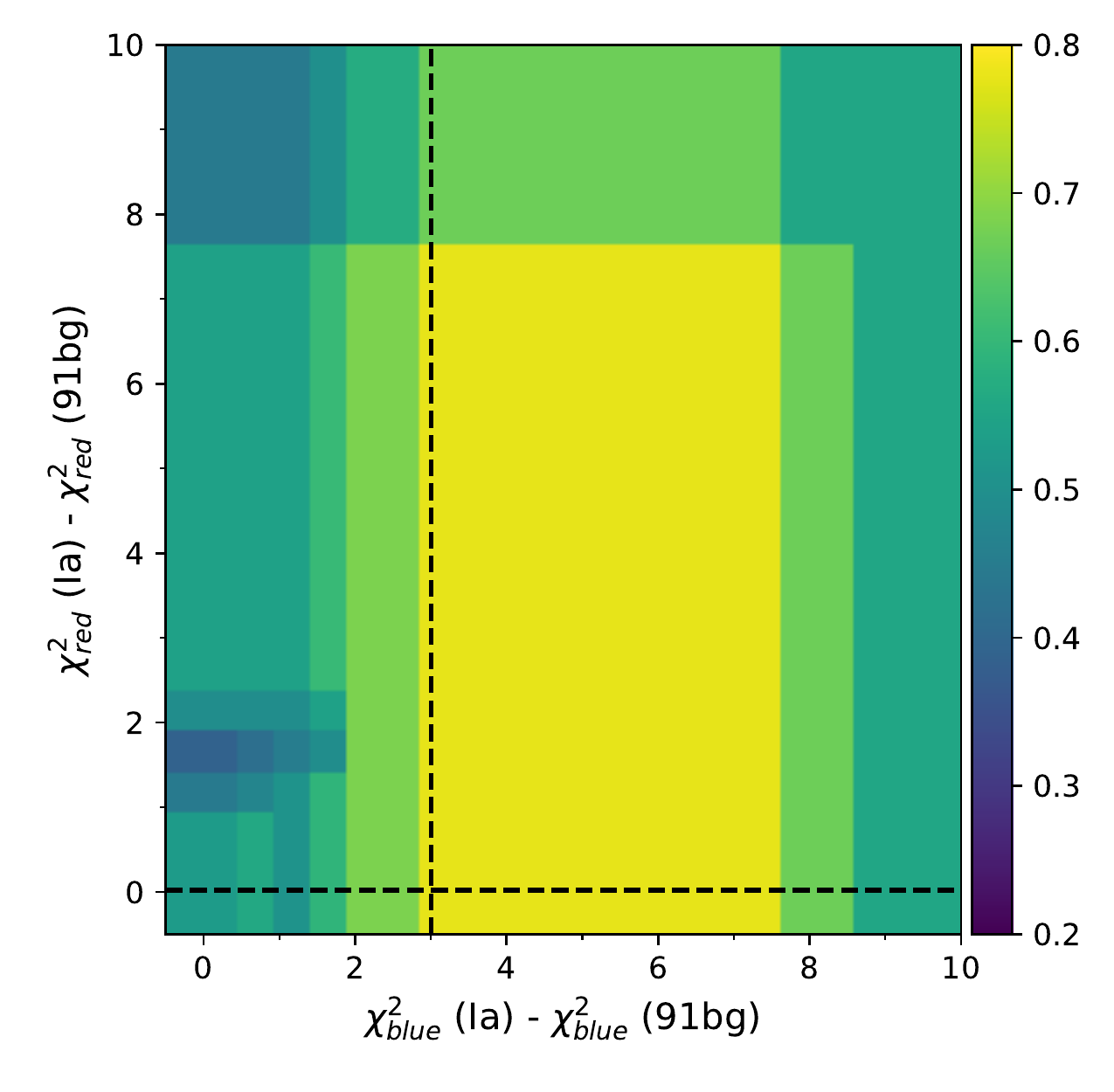}}

\caption{FOM values are shown as a function of classification boundaries. The higher dispersion of points when fitting red / blue observations as collective sets results in a significant area of degenerate FOM values that can only be broken by the inclusion of additional spectroscopically classified SNe Ia.}
\label{fig:fom_degeneracy}
\end{figure}

Following the prescription of G14, the rescaled CSP coordinates are used to optimize the value of Equation \ref{eq:fom} using a bootstrap technique \citep{Efron79, Felsenstein85}. In total, we draw 100 random samples, each containing 75\% of the available classification coordinates, and recalculate the FOM each time. So that the FOM can be evaluated, each sample is guaranteed to contain at least one 91bg-like SN. Shown in Figure \ref{fig:fom_coords}, the final boundaries are chosen using the average over all randomly realized samples. This results in a peak FOM value of 0.78 when fitting observations as red / blue sets and 0.75 when fitting bandpasses independently.

Shown in Figure \ref{fig:fom_degeneracy}, we note there exists a significant degeneracy in the maximized FOM value. This makes it possible to shift the classification boundaries in such a way that the classification of some targets changes despite the FOM remaining constant. To address this, we choose to use classification boundaries having the largest FOM while also independently minimizing the x and y cutoffs to be as close to the origin (0, 0) as possible. This results in classification boundaries of (3, 0) when fitting bandpasses as sets and (0.5, 0) when fitting bandpasses independently.

\subsubsection{Classification Results}
\label{ssec:classification_results}

\begin{figure}
\centering

\includegraphics[width=\columnwidth]{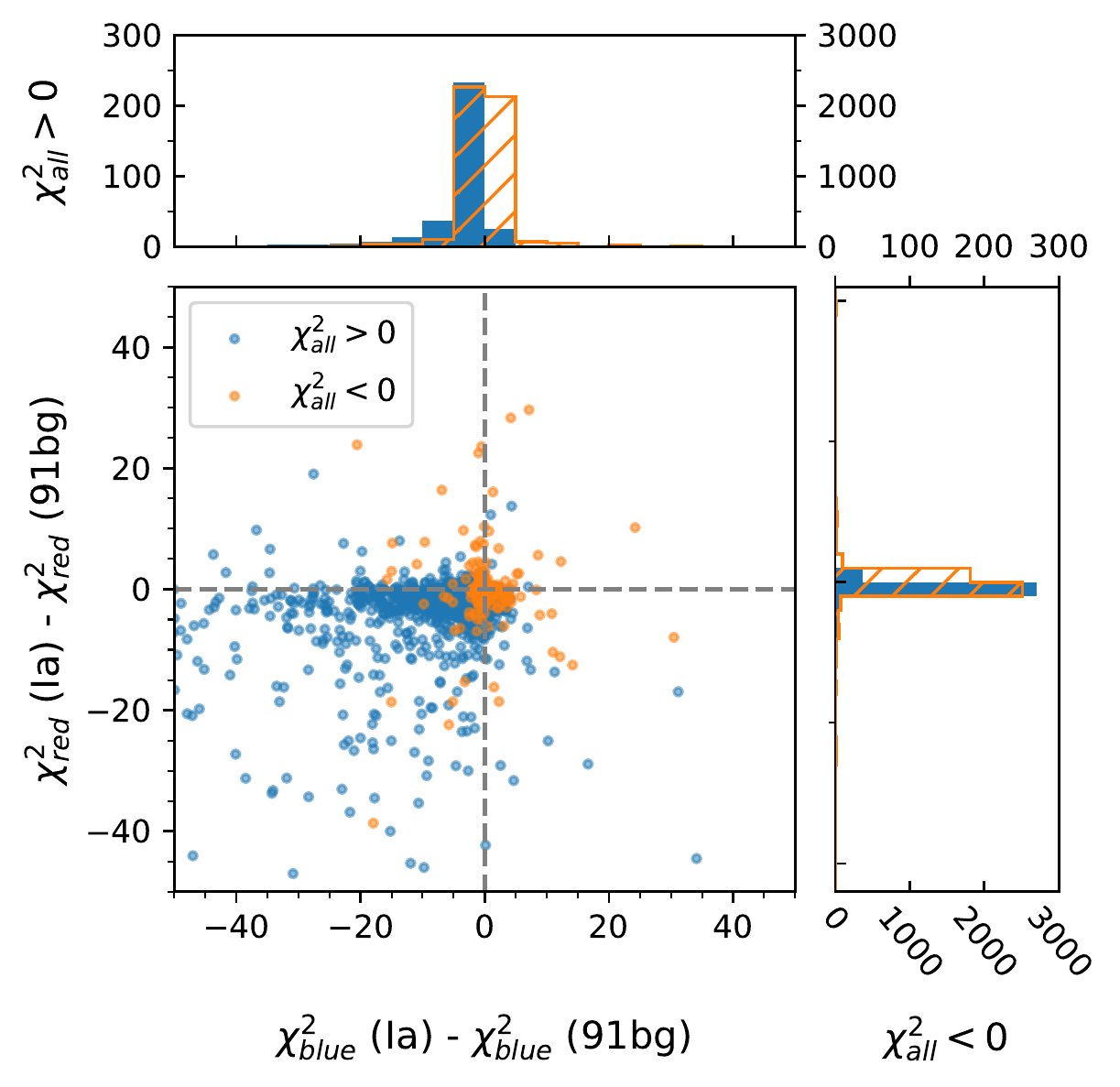}

\caption{Differences between the reduced $\chi^2$ of models for normal and SN~1991bg-like SNe. Fits are performed separately in the rest-frame blue ($\lambda_{\rm z, eff} < 5,500$ \AA) and red ($\lambda_{\rm z, eff} > 5,500$ \AA) bandpasses using a collective set of parameters for the redder and bluer bandpasses. We expect SN~1991bg-like objects to fall in the upper right quadrant (Q1) and normal SNe~Ia in the lower left (Q3). Light-curves with better overall fits (smaller  $\chi^2$) to all the data with the H07 (SN~1991bg) model are shown in blue (orange). 
}
\label{fig:classification}
\end{figure}

\begin{figure}
  \centering
  \includegraphics[width=\columnwidth]{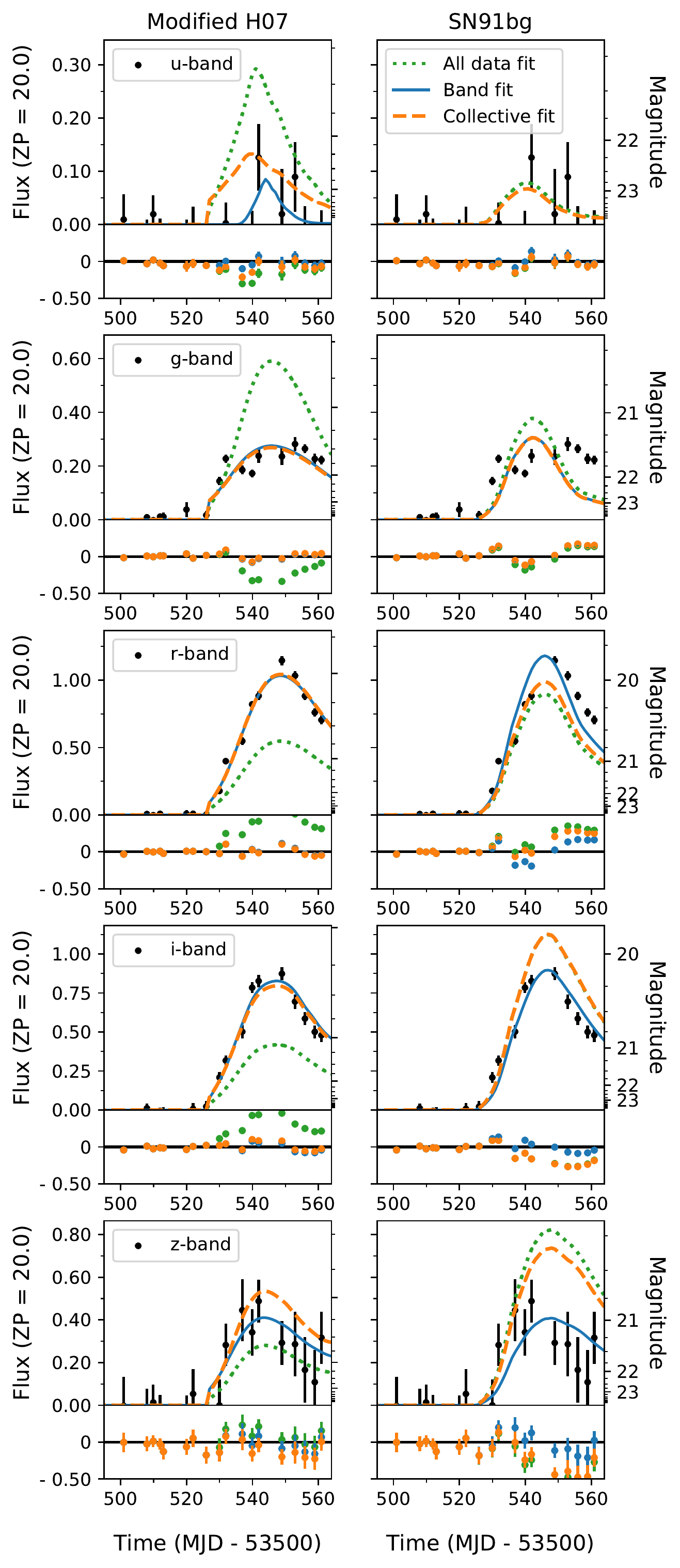}
  \caption{A comparison of fits to SDSS object CID 15749 using models for normal (left) and SN~1991bg-like (right) supernovae. Fits are performed to the entire data set (dotted green) the blue and red bandpasses as separate sets (dashed orange), and top each bandpass independently (solid blue). By fitting each model to subsets of the data, the impact of the observed color in constraining the fit is lessened and the morphology each band allowed to play a more influential role.
  }
  \label{fig:fit_comparison}
\end{figure}

\begin{figure*}
  \centering
  \includegraphics[width=.75\textwidth]{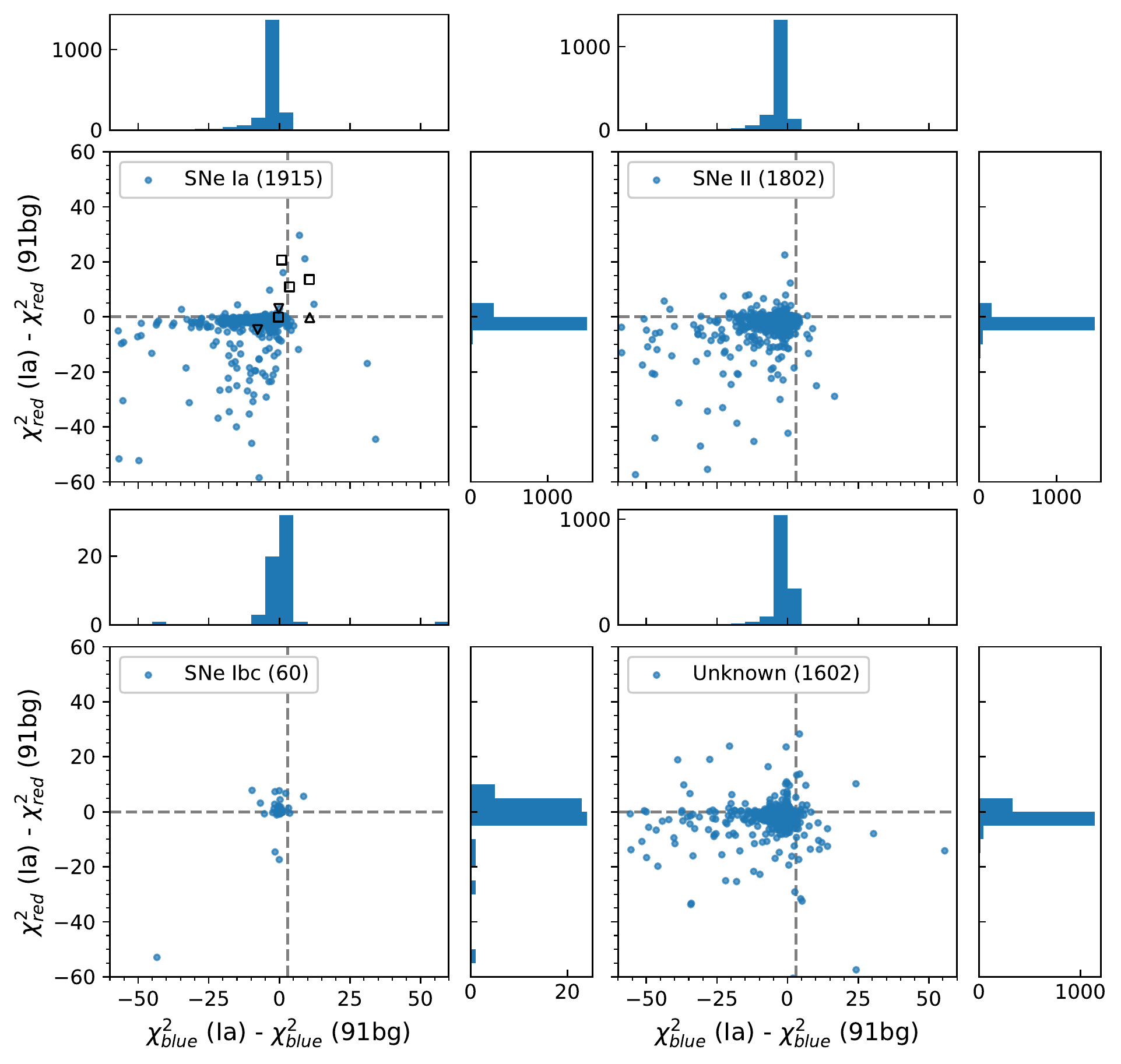}
  \caption{SNe observed by SDSS are broken down into four categories determined by their spectroscopic classification. If a spectroscopic classification is not available, the photometric classification determined by the PSNID software is used instead. The difference in reduced $\chi^2$ for models of normal and SN~1991bg-like SNe are shown for normal SNe~Ia (top left), SNe II (top right), SNe Ib/Ic (bottom left), and targets with light-curves too noisy to determine a classification (bottom right). We note that SN II events are constrained to quadrants 2 through 4 while SN Ib and Ic-like objects are clustered near the center of the phase space and SNe~Ia are primarily scattered across quadrants 1 and 3. Dashed lines are used to indicate $x=3$, $y=0$.}
  \label{fig:sub_classification}
\end{figure*}

Figure \ref{fig:fom_coords} shows the classification coordinates resulting from fitting bands independently and as collective red / blue sets. We see that when fitting the red and blue bands as collective sets, fits to red bandpasses have a stronger impact on the classification. This is demonstrated by the increased vertical dispersion of points where $y > 0$. Similarly, when fitting bandpasses independently the classification of SN~1991bg-like objects is primarily driven by poor template fits in the red bands. We attribute this to differences in the light-curve morphology.

We note that fits to the data as collective red / blue sets require fewer overall light-curve fits than when fitting bandpasses independently. Fitting the data as two sets only requires fitting each model twice, whereas fitting individual bands requires a number of fits equivalent to twice the number of observed bands.  As a result, fitting the observations as collective sets introduces fewer opportunities for a fit to diverge and the resulting classification coordinates are more stable. Collective fitting results also have a higher optimized FOM value. For these reasons, we choose to use coordinates determined from the collective fitting process to classify 91bg-like SNe. This leaves us with 16 remaining objects: CID 2778, 11570, 12689, 15204, 16215, 16309, 16692, 17094, 17468, 17886, 18218, 18751, 18890, 19065, 21678, 21898. Classification coordinates for these objects are listed in Table \ref{tab:classification_coords}.

\begin{figure*}
  \centering
  \includegraphics[width=\textwidth]{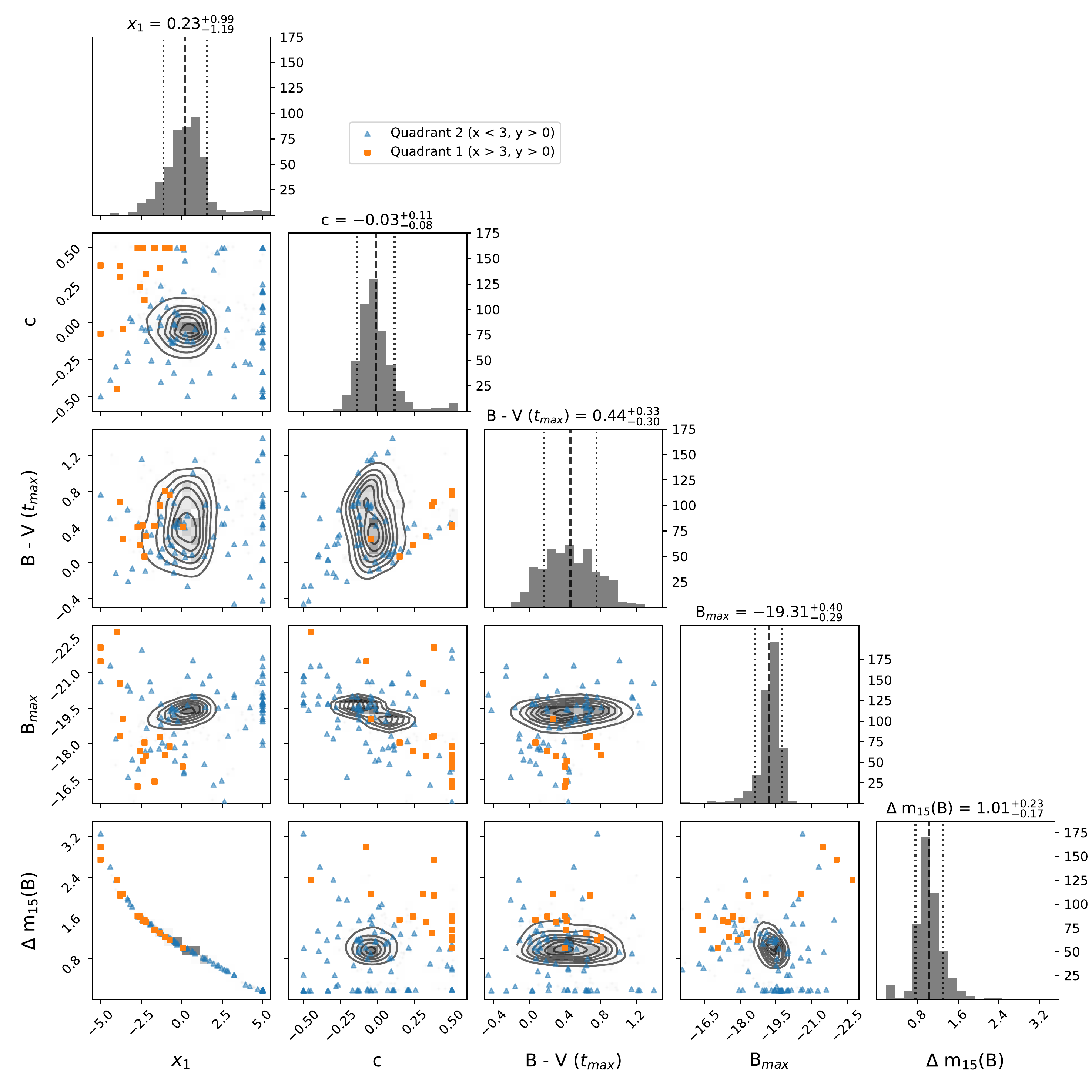}
  \caption{Distributions of the SALT 2.4 light-curve model fit to SDSS photometric observations. Points are color coded according to their position in the phase space $x = \chi^2_\text{blue} (\text{Ia}) - \chi^2_\text{blue} (\text{91bg})$, $y = \chi^2_\text{red} (\text{Ia}) - \chi^2_\text{red}(\text{91bg})$. We note that that SN~1991bg-like objects in quadrant 1 of the $x$, $y$ phase space (Q1; orange squares) are fainter and redder than the normal SNe~Ia population in quadrant 3 (Q1; grey density plot). Objects in quadrant 2 (Q2; blue triangles) are expected to be non-1991bg-like peculiar SNe.
  }
  \label{fig:salt2_params}
\end{figure*}

Sown in Figure \ref{fig:classification}, we note that some targets are classified as normal SNe~Ia in the red vs. blue $\chi^2_\nu$ comparison despite the SN~1991bg model having a lower $\chi^2_\nu$ when fit to all available data. An example case is shown in Figure \ref{fig:fit_comparison}, where we examine fits to observations of SDSS object CID 15749. In this case, we see that significant influence is exerted by poor fits to observations in \textit{ug} bands. However, visual inspection of the light-curve shows that the morphology of these bands is unusual despite normal behavior in the other bands. By separating the fits to the blue and red data, the impact of the $ug$ bands is mitigated, and the morphology of the other bands is allowed to play a more influential role in the classification.

To further understand the behavior of the employed classification scheme, in Figure \ref{fig:sub_classification} we compare our classification results against spectroscopic and photometric classifications from S18. We note that two of the four SNe with light-curves that were visually flagged by the SDSS SN team as potential 91bg-like objects are also labeled as 91bg-like SNe by our classifier. We also note that targets classified in the original data release as Type II SNe (SNe II), either spectroscopically or photometrically, are primarily constrained to quadrants two and three. The same can not be said for other types of core-collapse (CC) events, however both Type Ib and Ic SNe tend to be clustered in the center of the phase space. By asserting a cut at $x \ge 3$, $y \ge 0$, we find that all but one SNe II event can be excluded from the first quadrant (CID 5052). This indicates that, given the models we have chosen and their implementation, the classification scheme is robust against contamination by CC events.

Given that SN~1991bg-like SNe are redder at peak than their normal SNe~Ia counterparts \citep{Jaeger18, Galbany16}, it is expected for them to be more easily mistaken as CC explosions. It is thus surprising that so few CC SNe are classified as SN~1991bg-like SNe. In G14, contamination by CC SNe was explored using observations of 64 CC SNe from \cite{Anderson14} and other sources in the literature. Using a combination of cuts on magnitude, color, and the quality of fit in various bands, all but one SN were successfully removed. 

\begin{figure*}
\centering  

\gridline{
    \fig{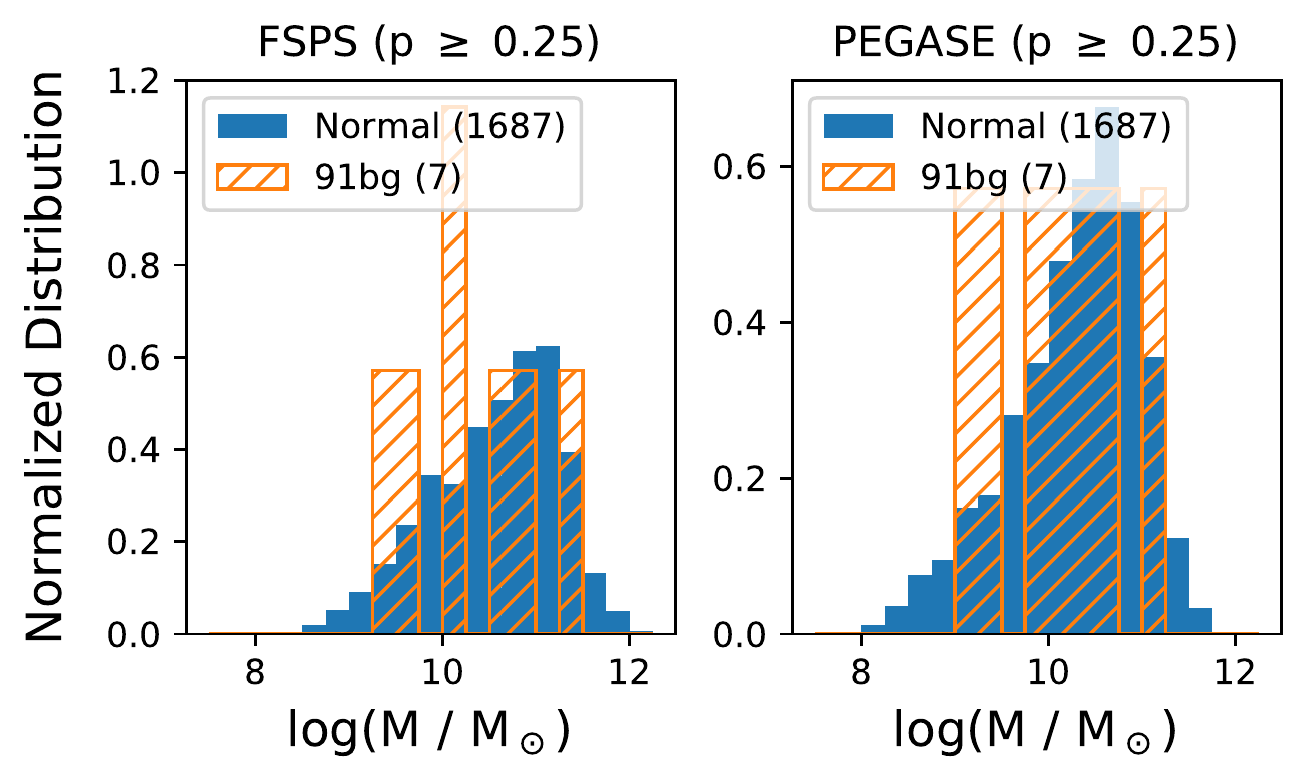}{\columnwidth}{(a) Mass estimates for SNe host galaxies as calculated by FSPS (left) and PEGASE (right).}
    \fig{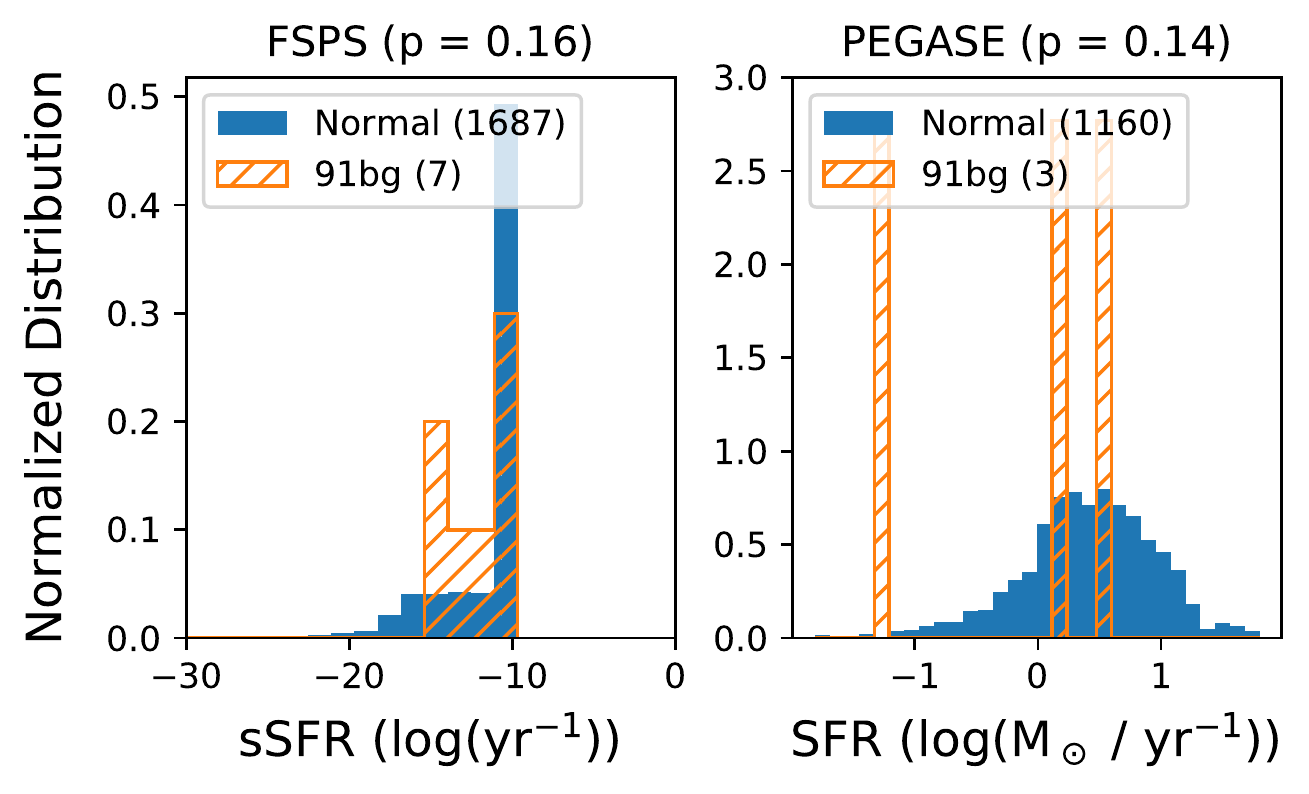}{\columnwidth}{(b) Star formation rates for SNe host galaxies as calculated by FSPS (left) and PEGASE (right). We note the number of targets with available sSFR estimates is different for each distribution.}
} 
\gridline{
    \fig{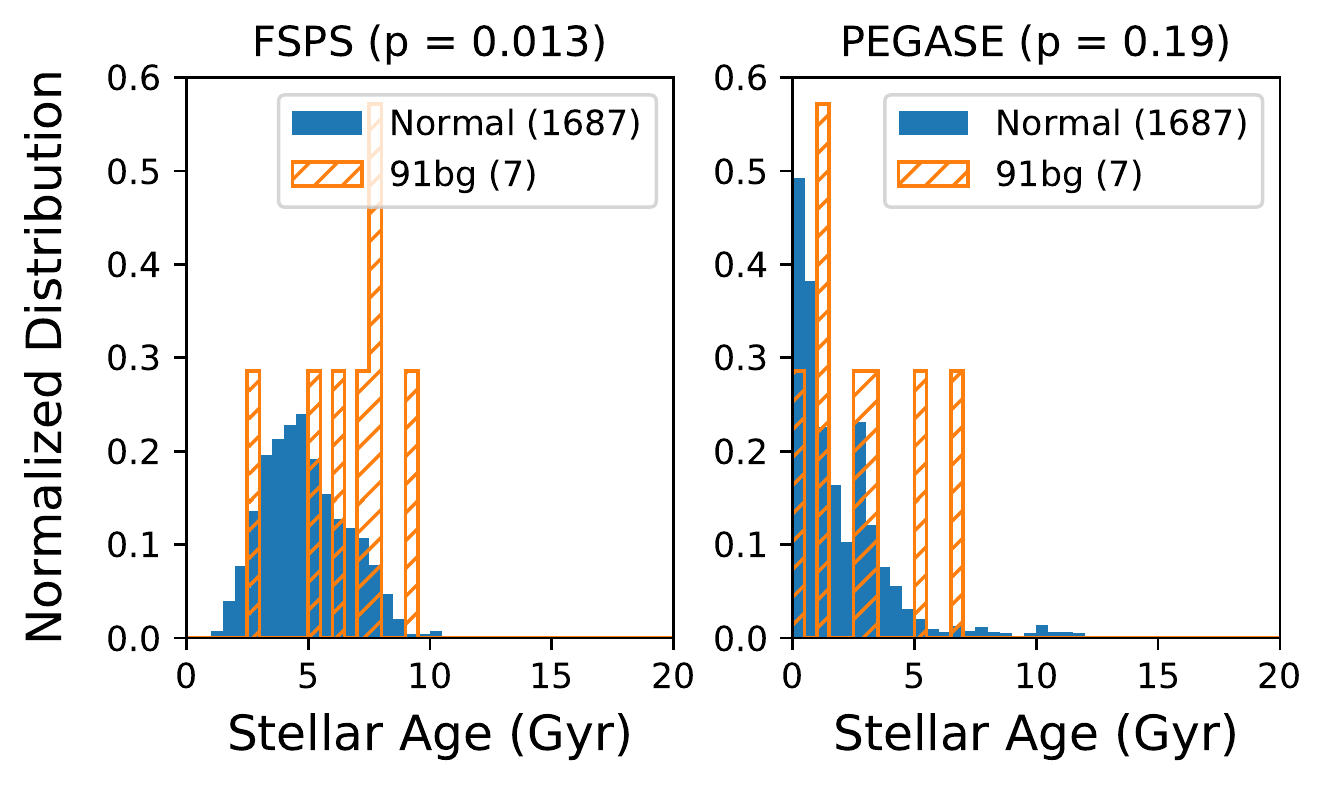}{\columnwidth}{(c) Average stellar age for SNe host galaxies as calculated by FSPS (left) and PEGASE (right).}
    \fig{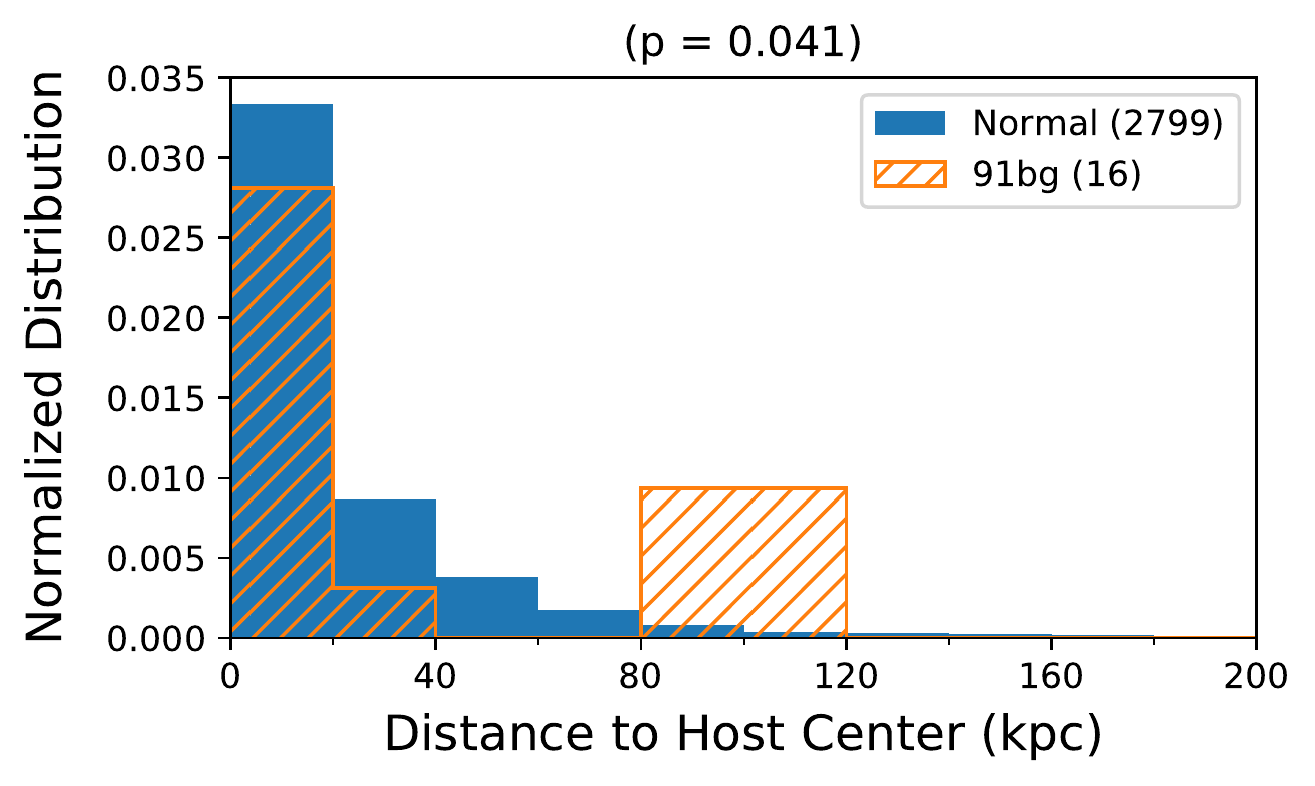}{\columnwidth}{(d) Physical distance of SNe from the photometric center of their host galaxies.}
} 
\caption{Area normalized distributions for host galaxy properties of objects identified as normal (solid blue) and SN~1991bg-like (dashed orange) SNe. The Aderson-Darling test is used to determine whether the two distributions are drawn from different underlying distributions and the resulting $p$-value is displayed for each host property. For $p$-values above the $5\%$ threshold, we conclude SN~1991bg-like events are drawn from the same underlying distribution. We find that 91bg-like events prefer further distances from the centers of their host galaxies and are more likely to occur in hosts with an older average stellar age than normal SNe~Ia.}
\label{fig:host_properties}
\end{figure*}

Our implementation of the classification procedure displays a similar level of resistance to CC contamination as the original implementation of G14. However, in this work we are able to exclude CC events without the need for additional cuts on individual SN properties. One possible explanation is a significant intrinsic bias in the types of targets observed by the SDSS survey. Another possibility are differences in the models chosen for normal/SN~1991bg-like events, but this is minimized since we have specifically chosen models similar to those used in G14. 

We thus conclude that this change in behavior is primarily driven by changes we have imposed in the way parameters are varied to fit each model.  By choosing to vary parameters either independently across bands or as collective red / blue sets we have chosen to favor either light-curve morphology or color in the fitting process. In comparison, the way in which SiFTO varies parameters across bands while enforcing inter-bandpass relationships provides a middle-ground between these approaches.  SiFTO also relies on significantly more free parameters, thus improving the quality of the overall fit to non-SNe Ia.

Although the SDSS data release did not include a dedicated SN sub-typing effort, objects classified by SDSS were submitted to the Open Supernova Catalog \citep{Guillochon17}, and that collection was collectively analyzed by \citet{Pruzhinskaya19} for anomalous light-curves using a random-forest machine learning classifier. Listed in Table \ref{tab:osc_comparison}, a total of 37 SN from the SDSS sample were identified as peculiar objects. Out of these, none of these targets are included in our selection of 91bg-like SNe.

We note that the number of selected targets is significantly less than what is expected according to the rate of 6 -- 15 \% claimed in the literature \citep{Ganeshalingam10, Li11, Gonzalez_Gaitan11, Silverman12}. However, we expect to observe fewer SNe than the predicted rates due to intrinsic survey bias towards the identification of normal (brighter) SNe~Ia.  Additionally, the faintness and narrowness of SN~1991bg-like light-curves mean they spend less time over the SNR = 5 limit. This makes them less likely to be selected for followup and also more likely to be removed by quality cuts.

\subsection{Properties of Selected SNe} \label{ssec:properties_of_91bg}

\begin{figure*}

\gridline{
    \fig{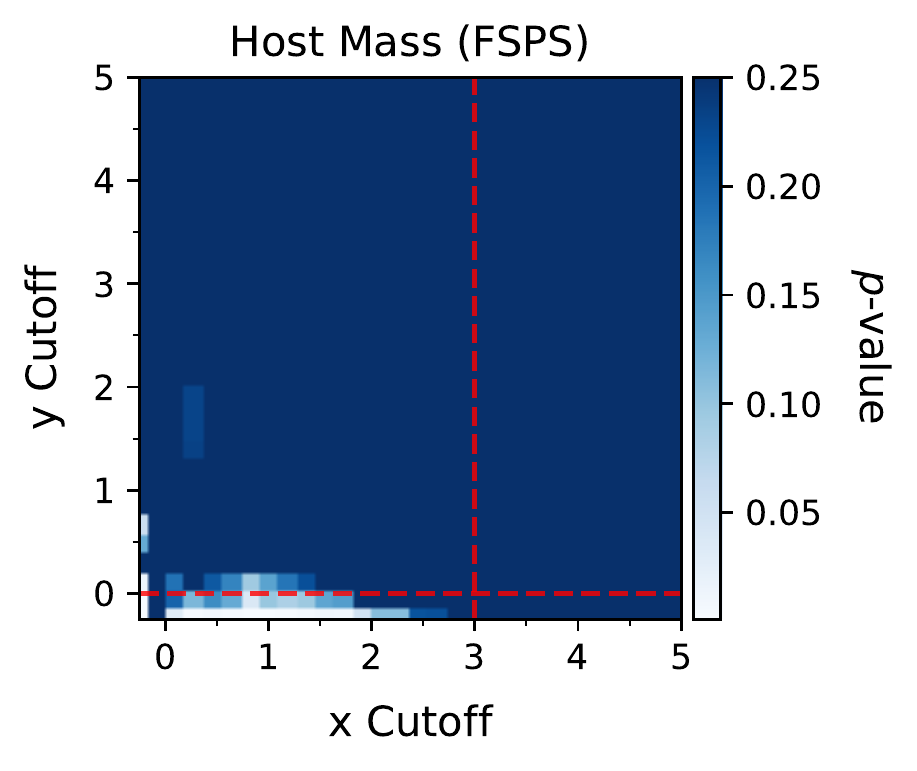}{.33\textwidth}{}
    \fig{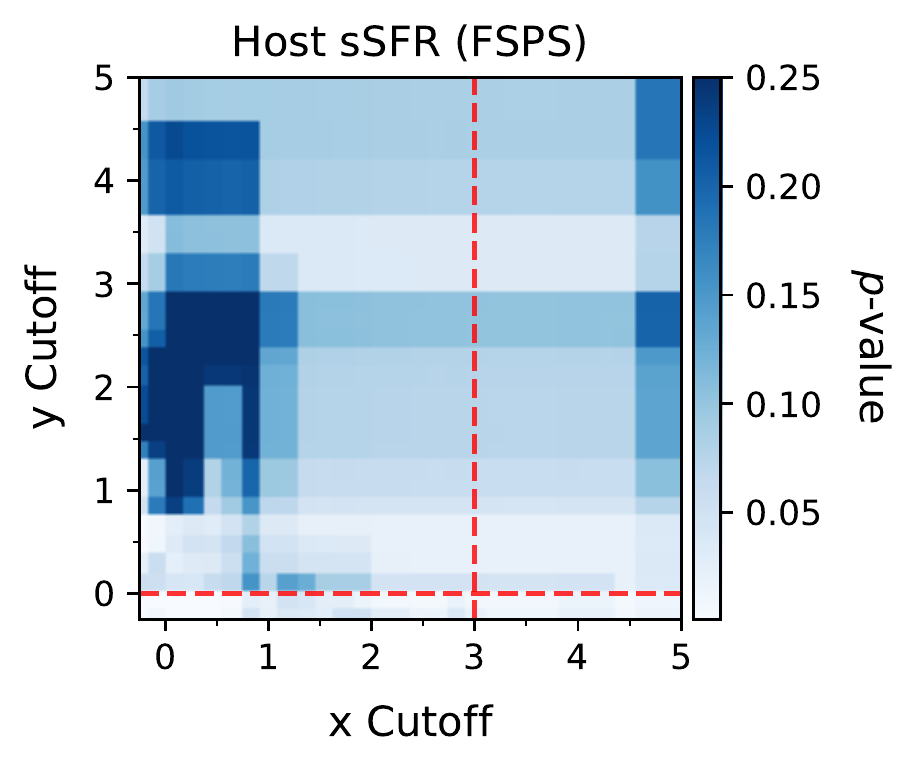}{.33\textwidth}{}
    \fig{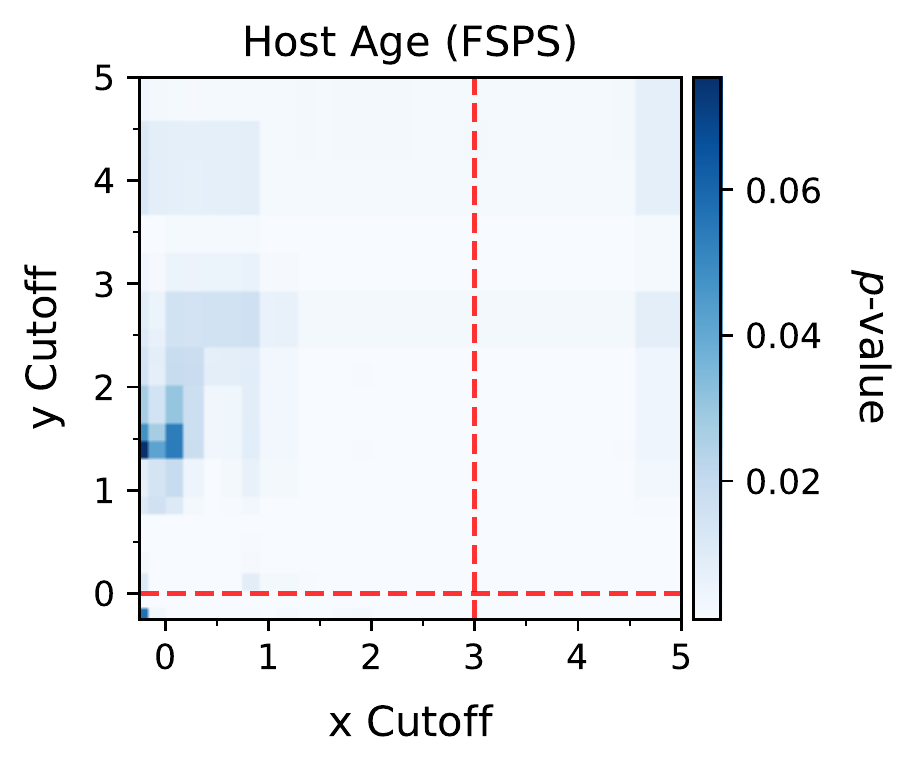}{.33\textwidth}{}
} 

\gridline{
    \fig{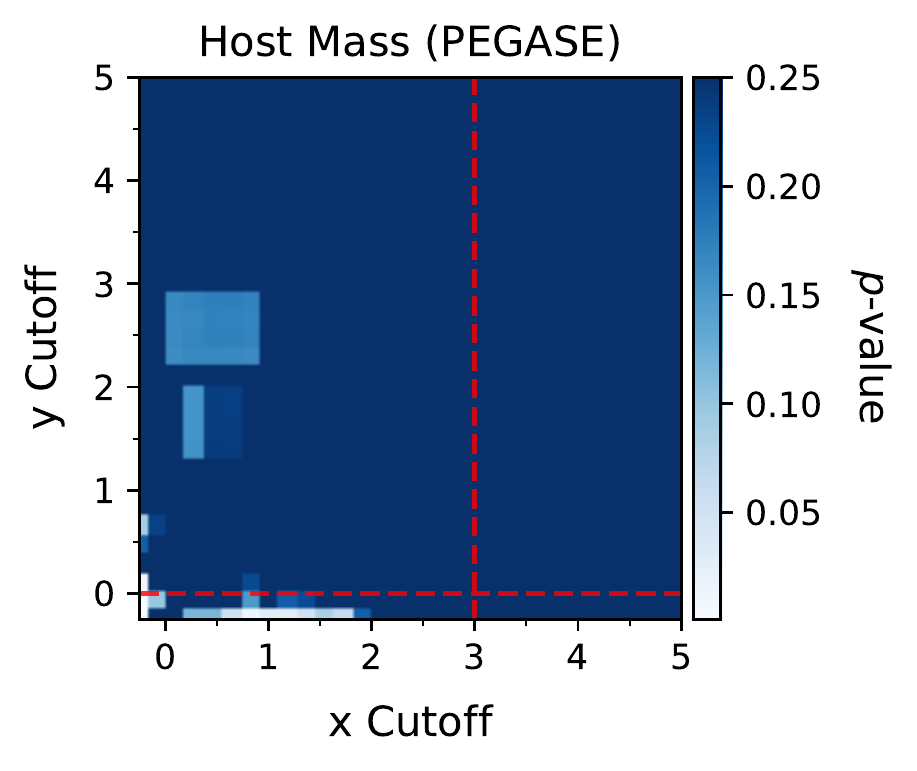}{.33\textwidth}{}
    \fig{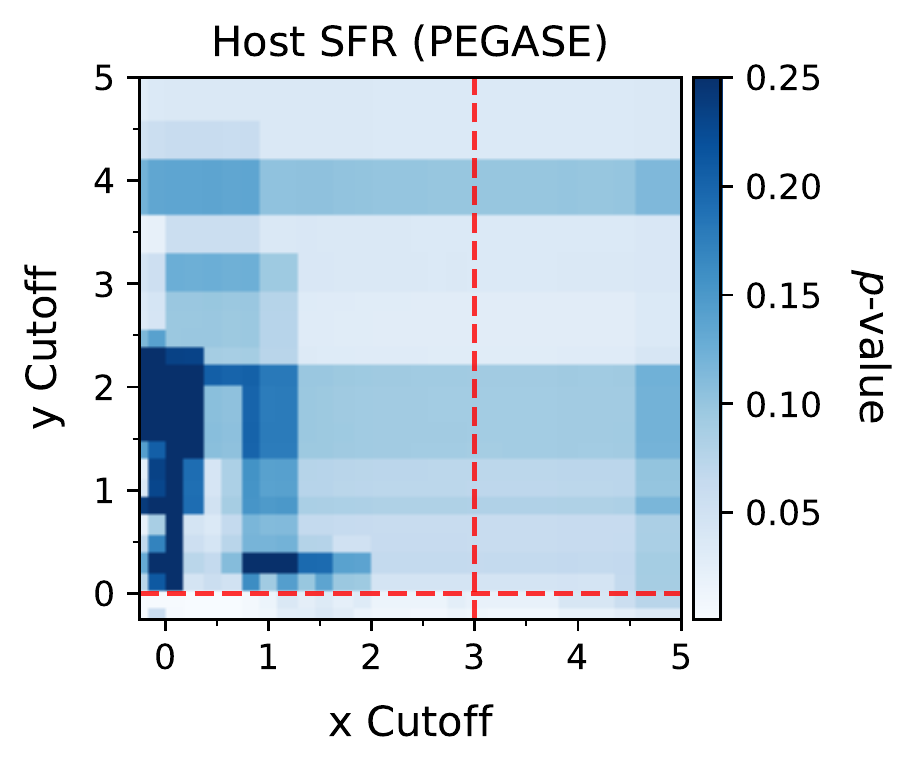}{.33\textwidth}{}
    \fig{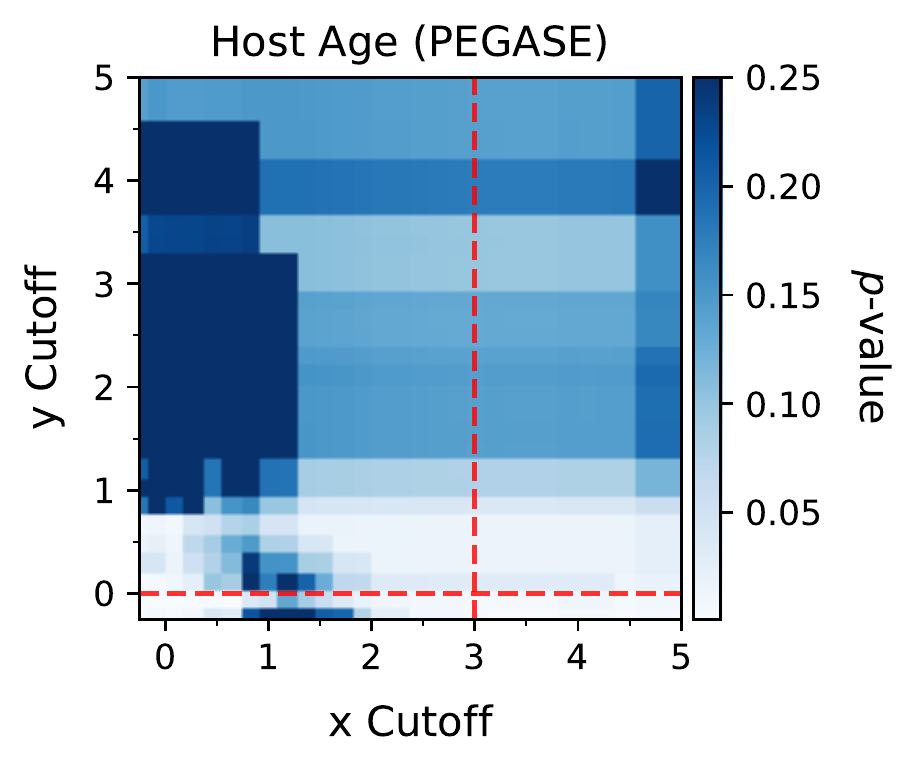}{.33\textwidth}{}
} 

\gridline{
    \fig{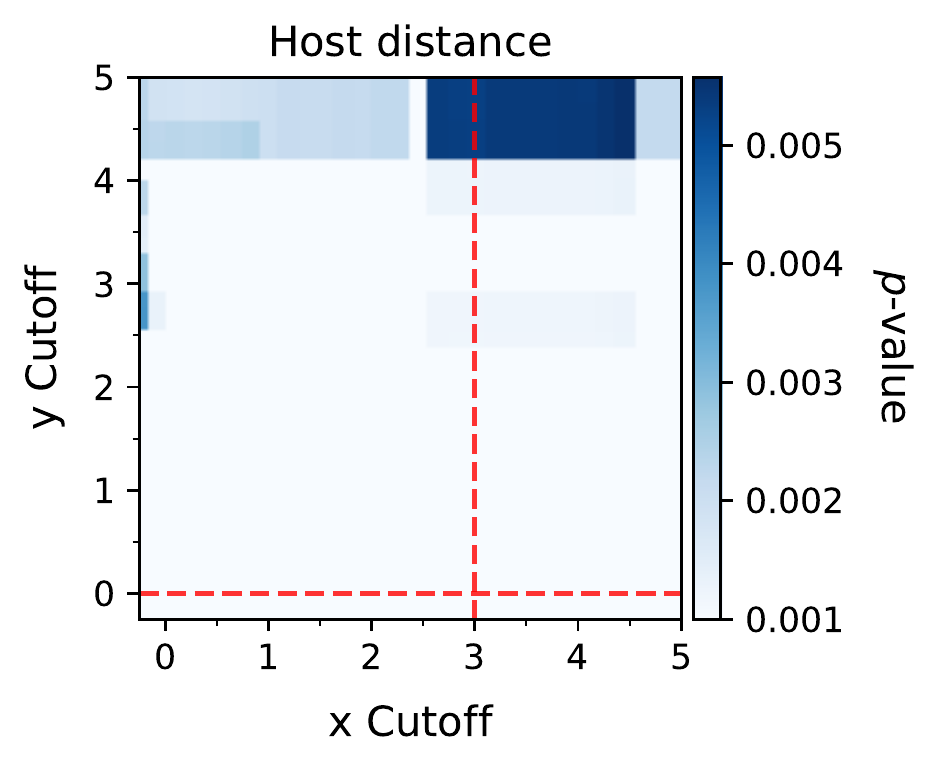}{.33\textwidth}{}
} 

\caption{The Anderson-Darling test is used to determine whether normal and 91bg-like SNe are drawn from the same underlying distributions of host galaxy properties. This test is performed as a function of quadrant boundaries for the classification coordinates. The resulting $p$-values are shown for a collection of host galaxy properties as calculated by FSPS (top row) and PEGASE (middle row). Considered properties in these rows include host galaxy mass (left), star formation rate (center) and average stellar age (right). The calculation is also repeated for the physical distance of SNe from their host galaxies (bottom row). The actual quadrant boundaries used in this work are shown in red for reference. The $p$ values are not qualitatively sensitive to the choice of quadrant boundary --- for host mass, average stellar age, and distance the $p$ values consistently fall either above or below $0.05$. 
The normal SNe~Ia and 91bg-like SNe~Ia identified in this work are clearly different in their host galaxy stellar age and physical distance from the host galaxy center.
}
\label{fig:pvalue}
\end{figure*}

To understand the intrinsic behavior of our selected SNe, we fit each target with SALT 2.4 and list the results in Table \ref{tab:salt2_fits}. Figure \ref{fig:salt2_params} shows that objects identified by our classifier follow many of the trends we expect of 91bg-like objects. In terms of color, selected objects tend to be redder than normal SNe with an average \textit{B - V} color of $0.88$ mag. They also fall on the dimmer and faster-declining extremes with an average $\Delta \text{m}(B)_{15}$ of  $1.56$.  We note that SNe selected in quadrant 2 are much more diverse in their parameter distribution, but still have a notable subset of SNe lying on the faint and fast extremes of the SNe population. 

%   \centering
%   \includegraphics[width=.9\columnwidth]{ssfr_pvalue.pdf}
%   \caption{The  Anderson-Darling test is used to determine whether normal and 91bg-like SNe are drawn from the same underlying distributions of host galaxy properties. This test is performed as a function of quadrant boundaries for the classification coordinates. The resulting $p$-values are shown for host galaxy mass (top) and distance of the SN from the host galaxy center. $p$-values are bounded to a minimum of $0.001$ and a maximum of $0.25$. Boundaries chosen for use in this work are marked as red dashed lines. We note the difference in scale between the top two and the bottom panel.}
% \end{figure}term

Figure \ref{fig:host_properties} shows the relationship between SNe and their host galaxies using properties determined in S18 with the FSPS \citep{Conroy09, Conroy10} and PEGASE \citep{Fioc97} software routines. To understand whether the selected SNe are drawn from the same underlying host galaxy distribution as normal SNe, we perform an Anderson-Darling test \citep{Anderson52}. We take as a null-hypothesis that the two populations are drawn from the same underlying distribution. For a $p$-value $<5\%$ we reject the null hypothesis and assert that the underlying distributions are different. In all cases presented by this work, $p$-values are bound to the range $0.001 \leq p \leq 0.25$. 

We find no statistical evidence to indicate the selected SNe are drawn from a different underlying distribution of galaxy mass or Star Formation Rate (SFR). Although the $p$-value determined for SFR using FSPS and PEGASE differ significantly ($24\%$ and $16\%$ respectively), this can be attributed to the fact that the FSPS routine was able to determine SFR values for more 91bg-like targets than the PEGASE routine. In practice, the number of available points is considered in the calculation of the $p$-value \citep[see][]{Scholz87}. However, we also cannot rule out the possibility that the subsample enforced by PEGASE is somehow biased towards a particular distribution of SFR. 

We find the only considered properties to indicate a different underlying distribution is the average stellar age and distance from the center of the host galaxy. Visual inspection of Figure \ref{fig:host_properties} shows that objects identified by our classifier as being 91bg-like prefer galaxies with older stellar populations and have a higher probability of occurring further away from the center of the galaxy.  This confirms a previous result found using SDSS data in \citet[][hereafter G12]{Galbany12}. Using a subset of 200 spectroscopically or photometrically confirmed SNe~Ia at redshifts $z \leq 0.25$, G12 found that the average fitted color term ($c$) from SALT decreased with the projected distance for SNe~Ia in spiral galaxies. It was also determined that SNe in elliptical galaxies tend to have narrower light-curves if they explode at larger distances, although the impact of selection effects was unclear.

\begin{figure}
  \includegraphics[width=\columnwidth]{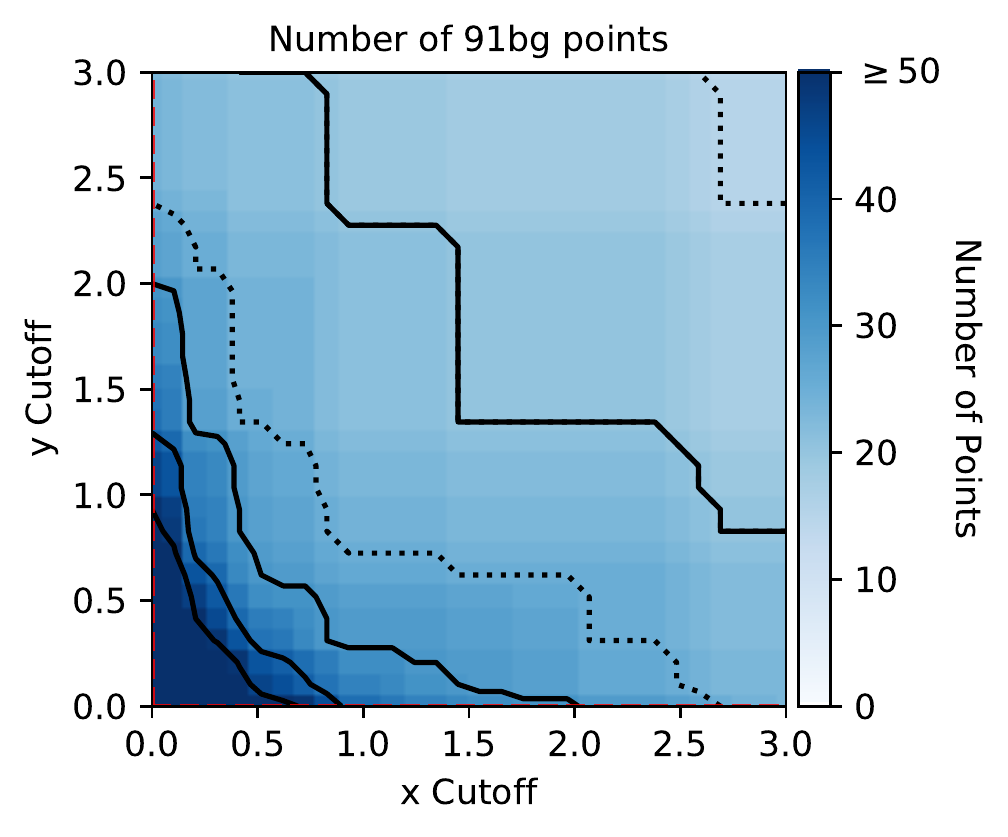}
  \caption{The number of objects classified as SN~1991bg-like as a function of quadrant boundaries for the classification coordinates. Contours are shown in steps of 10 (solid) and 5 (dashed) SNe.}
  \label{fig:num_points}
\end{figure}

In principle, the results of the Anderson-Darling test are dependent on the quadrant boundaries used to classify targets. Figure \ref{fig:pvalue} shows the recalculated $p$-values for a range of classification boundaries. We see that for targets that have passed our quality cuts, there is minimal variation in $p$-values surrounding our chosen quadrant boundaries for host galaxy mass, age, SFR, and distance. For an $x$ and $y$ cutoff large enough, we do see a slight increase in the $p$-value for some properties. However, as shown in Figure \ref{fig:num_points}, there are only a small number of points selected at those extremes and the existence of additional selection effects becomes unclear.

\section{Conclusion} \label{sec:conclusion}

Using SN observations from SDSS, we explore the implementation of an empirically based classification technique targeted at the identification of SN~1991bg-like SNe. In the presented approach dedicated light-curve fits are performed for observational data in rest-frame blue ($\lambda_{\rm z, eff} < 5,500$  \AA) and red ($\lambda_{\rm z, eff} > 5,500$ \AA) bandpasses. Using models for both a normal and 91bg-like SN, targets are classified based on the difference in reduced $\chi^2$ values for each model in blue and red wavelengths.

We consider two distinct implementations of this technique. In the first implementation each observed bandpass is fitted independently and the $\chi^2$ values from each fit are summed to determine the overall $\chi^2$ for the blue and red bandpasses respectively. The second implementation is performed in the reverse order: Observed bandpasses are split into sets of bluer and redder data and then fit as two collective sets. We find no significant differences in the classifications generated from either approach. However, we note that the latter approach requires a larger number of spectroscopically caliisifed SNe Ia to fully train the classification procedure.

To understand the potential for contamination by non-SNe~Ia, we compare our classification results with spectroscopic classifications for a limited subset of SDSS targets. We find that our classification procedure is robust against contamination from core collapse events with only one SNe II being classified as SN~1991bg-like. When performing the same comparison against photometric classifications from PSNID, we reach the same conclusion. 

In total our classifier identifies 16 SNe from the SDSS-II SN sample: CID 2778, 11570, 12689, 15204, 16215, 16309, 16692, 17094, 17468, 17886, 18218, 18751, 18890, 19065, 21678, 21898.
Existing subtypes for SDSS SNe in the literature is limited, restricting our ability to compare results. A total of 37 targets from the SDSS sample were identified as anomalous objects in an external analysis using a random-forest classifier on photometric data from the OSC. Out of these objects, we classify none of them as SN~1991bg-like events.  

Using host galaxy properties from SDSS, we investigate potential differences in the distribution of host galaxy properties for normal SNe and those selected by our classifier. An inspection of host mass as measured by the FSPS and PEGASE routine reveals no statistically significant bias in host galaxy mass. We also find no significant trend in host galaxy SFR. However, selected objects are seen to prefer galaxies with older stellar populations and have a higher probability of occurring further away from the center of the galaxy.

Future work is currently planned to extend the presented classification technique to classify other kinds of peculiar SNe. A more detailed investigation is also planned to explore potential biases that may be introduced in the way spectroscopic templates are varied to fit photometric observations. 
\acknowledgments

D.J.P., M.W.-V., and L.G. were supported in part by the U.S. Department of Energy, Office of Science, and Office of High Energy Physics under award number DE-SC0007914. L.G. was also funded by the European Union’s Horizon 2020 research and innovation program under the Marie Sk\l{}odowska-Curie grant agreement No. 839090. S.G.-G. was supported by FCT under Project CRISP PTDC/FIS-AST-31546. We also thank Ella Kane and Anish Bhagwat for their assistance in visually inspecting fits of multiple models to photometric data.

\facilities{Sloan, LSST}
\software{
    AstroPy~\citep{Astropy18}\footnote{\url{http://www.astropy.org}},
    NumPy\footnote{\url{http://www.numpy.org}}, 
    SciPy\footnote{\url{http://www.scipy.org}},
    SNData \footnote{\url{https://sndata.readthedocs.io/}},
    SNCosmo~\citep{Barbary16}.\footnote{\url{https://sncosmo.readthedocs.io/}}
}

\clearpage
\bibliographystyle{aasjournal}
\bibliography{main.bib}

\begin{deluxetable}{cc}
\tablecaption{Objects identified in the SDSS SN Survey as being potentially peculiar objects based on visual inspection of the spectrophotometric properties. Objects are listed using their Candidate Identifier from \citet{Sako18}} \label{tab:sdss_peculiars}
\tablehead{\colhead{CID} & \colhead{Classification}}
\startdata
4524 & SN 2002ci \\
6295 & SN 1991bg \\
7017 & SN 2002ci \\
8151 & SN 2002cx \\
12979 & SN 1991bg \\
13357 & SN 2002cx \\
15340 & SN 2000cx \\
17886 & SN 1991bg \\
18890 & SN 1991bg \\
20208 & SN 2002cx
\enddata
\end{deluxetable}

\begin{longrotatetable}
\begin{deluxetable}{rllllrrrrrrrrrrr}
\tablecaption{Fitted parameters are listed for a combination of models and bandpass collections. \textit{red} and \textit{blue} bands indicate a collections bandpasses where the rest-frame effective wavelength is redward or blueward of 5500 \AA. Fits to individual bands are listed using the banpass name and SDSS CCD column number (see \cite{Doi10}). We note that the version of the Hsiao model used does not include a color parameter $c$ and thus has one degree of freedom more than the SN 1991bg model. Any redshift values missing a reported error were specified using spectroscopic measurements. Any remaining missing entries for a particular fit indicate that parameter was not included in the given fit and the result from a fit to all data was used instead. Results are limited to the first 20 table entries. A full version of the table is available online.} \label{tab:class_fits}
\tablehead{\colhead{CID} & \colhead{band} & \colhead{model} & \colhead{z} & \colhead{z err} & \colhead{t0} & \colhead{t0 err} & \colhead{c} & \colhead{c err} & \colhead{x1} & \colhead{x1 err} & \colhead{E(B - V)} & \colhead{chisq} & \colhead{ndof} & \colhead{B$_{\rm max}$} & \colhead{$\Delta M(\text{B})_{15}$}\\ \colhead{ } & \colhead{ } & \colhead{ } & \colhead{ } & \colhead{ } & \colhead{MJD} & \colhead{MJD} & \colhead{ } & \colhead{ } & \colhead{ } & \colhead{ } & \colhead{$\mathrm{mag}$} & \colhead{ } & \colhead{ } & \colhead{$\mathrm{mag}$} & \colhead{$\mathrm{mag}$}}
\startdata
679 & all & Hsiao & 0.124 &  & 53690.37 & 0.01 &  &  & 0.50 & 0.72 & 0.07 & 68.10 & 51.0 & -15.80 & 0.52 \\
679 & all & sn91bg & 0.124 &  & 53689.20 & 0.05 & 0.00 & 0.78 & 1.25 & 0.06 & 0.07 & 81.61 & 50.0 & -15.72 & 1.45 \\
679 & blue & Hsiao & 0.124 &  & 53690.37 &  &  &  & 0.33 & 0.30 & 0.07 & 16.90 & 19.0 & -15.64 & 0.65 \\
679 & blue & sn91bg & 0.124 &  & 53689.20 &  & 0.00 & 0.92 & 1.25 & 0.00 & 0.07 & 22.08 & 18.0 & -16.14 & 1.45 \\
679 & g5 & Hsiao & 0.124 &  & 53690.37 &  &  &  & 0.30 & 0.31 & 0.07 & 7.75 & 9.0 & -15.63 & 0.68 \\
679 & g5 & sn91bg & 0.124 &  & 53689.20 &  & 0.00 & 0.63 & 1.25 & 0.00 & 0.07 & 12.13 & 8.0 & -16.12 & 1.45 \\
679 & i5 & Hsiao & 0.124 &  & 53690.37 &  &  &  & 0.50 & 0.15 & 0.07 & 20.59 & 9.0 & -16.54 & 0.52 \\
679 & i5 & sn91bg & 0.124 &  & 53689.20 &  & 0.27 & 0.31 & 1.25 & 0.00 & 0.07 & 28.69 & 8.0 & -15.18 & 1.45 \\
679 & r5 & Hsiao & 0.124 &  & 53690.37 &  &  &  & 0.50 & 0.17 & 0.07 & 11.83 & 9.0 & -15.98 & 0.52 \\
679 & r5 & sn91bg & 0.124 &  & 53689.20 &  & 0.03 & 0.55 & 1.25 & 0.00 & 0.07 & 17.34 & 8.0 & -15.67 & 1.45 \\
679 & red & Hsiao & 0.124 &  & 53690.37 &  &  &  & 0.50 & 0.10 & 0.07 & 46.75 & 31.0 & -16.14 & 0.52 \\
679 & red & sn91bg & 0.124 &  & 53689.20 &  & 0.00 & 0.62 & 1.25 & 0.00 & 0.07 & 57.97 & 30.0 & -15.62 & 1.45 \\
679 & u5 & Hsiao & 0.124 &  & 53690.37 &  &  &  & 0.50 & 0.85 & 0.07 & 8.12 & 8.0 & -16.62 & 0.52 \\
679 & u5 & sn91bg & 0.124 &  & 53689.20 &  & 0.00 & 0.50 & 1.25 & 0.03 & 0.07 & 9.33 & 7.0 & -17.52 & 1.45 \\
679 & z5 & Hsiao & 0.124 &  & 53690.37 &  &  &  & -0.09 & 0.26 & 0.07 & 10.95 & 9.0 & -17.13 & 1.22 \\
679 & z5 & sn91bg & 0.124 &  & 53689.20 &  & 0.30 & 0.65 & 1.15 & 0.05 & 0.07 & 11.80 & 8.0 & -15.07 & 1.52 \\
682 &  &  &  &  &  &  &  &  &  &  &  &  &  &  &  \\
685 & all & Hsiao & 0.205 & 0.012 & 53654.83 & 0.85 &  &  & 0.50 & 0.00 & 0.05 & 343.62 & 85.0 & -18.47 & 0.52 \\
685 & all & sn91bg & 0.050 & 0.019 & 53647.87 & 4.49 & 0.00 & 0.01 & 1.25 & 0.00 & 0.05 & 721.75 & 84.0 & -15.08 & 1.45 \\
685 & blue & Hsiao & 0.205 &  & 53654.83 &  &  &  & 0.50 & 0.00 & 0.05 & 259.79 & 51.0 & -18.45 & 0.52
\enddata
\end{deluxetable}
\end{longrotatetable}

\begin{longrotatetable}
\begin{deluxetable}{cccccccccccccc}
\tablecaption{Parameters are listed for fit results of the Salt 2.4 model to SDSS SNe. Any redshift values missing a reported error were specified using spectroscopic measurements. For each fit the stretch parameter $x_1$ was bounded to the interval [-5, 5] and the color parameter $c$ to the interval [-0.5, 0.5]. If a spectroscopic redshift was not available, the redshift was bounded to [0, 1] and fit for. Results are limited to the first 20 table entries. A full version of the table is available online.} \label{tab:salt2_fits}
\tablehead{\colhead{CID} & \colhead{z} & \colhead{z err} & \colhead{t0} & \colhead{t0 err} & \colhead{c} & \colhead{c err} & \colhead{x1} & \colhead{x1 err} & \colhead{E(B - V)} & \colhead{chisq} & \colhead{ndof} & \colhead{B$_{\rm max}$} & \colhead{$\Delta M(\text{B})_{15}$}\\ \colhead{ } & \colhead{ } & \colhead{ } & \colhead{MJD} & \colhead{MJD} & \colhead{ } & \colhead{ } & \colhead{ } & \colhead{ } & \colhead{Mag} & \colhead{ } & \colhead{ } & \colhead{$\mathrm{mag}$} & \colhead{$\mathrm{mag}$}}
\startdata
679 & 0.124 &  & 53689.20 & 1.25 & 0.50 & 0.19 & 5.00 & 9.46 & 0.07 & 62.20 & 43.0 & -15.45 & 0.19 \\
682 &  &  &  &  &  &  &  &  &  &  &  &  &  \\
685 & 0.327 & 0.016 & 53652.64 & 0.90 & 0.04 & 0.10 & 5.00 & 0.13 & 0.05 & 577.95 & 94.0 & -19.37 & 0.18 \\
688 & 0.067 &  & 53621.86 & 0.00 & 0.50 & 0.01 & 1.35 & 0.98 & 0.07 & 86.74 & 52.0 & -15.48 & 0.80 \\
691 & 0.130 &  & 53606.20 & 0.49 & 0.02 & 0.05 & -0.46 & 0.31 & 0.06 & 32.98 & 32.0 & -19.05 & 1.14 \\
694 & 0.126 &  & 53622.89 & 0.00 & 0.25 & 0.01 & 1.11 & 0.15 & 0.06 & 364.52 & 96.0 & -18.89 & 0.85 \\
696 & 0.549 & 0.040 & 53623.64 & 1.13 & -0.08 & 0.10 & -0.45 & 0.67 & 0.03 & 77.69 & 71.0 & -20.88 & 1.14 \\
697 & 0.155 &  & 53617.20 & 1.24 & 0.50 & 0.01 & 1.56 & 0.66 & 0.07 & 193.63 & 96.0 & -17.62 & 0.77 \\
701 & 0.205 &  & 53609.93 & 0.00 & 0.10 & 0.04 & -1.76 & 0.40 & 0.05 & 63.81 & 56.0 & -19.20 & 1.43 \\
703 & 0.296 &  & 53626.57 & 0.61 & -0.01 & 0.04 & 0.72 & 0.59 & 0.05 & 64.60 & 81.0 & -19.50 & 0.93 \\
704 & 0.205 &  & 53607.94 & 0.00 & 0.21 & 0.07 & -0.71 & 0.63 & 0.06 & 54.99 & 51.0 & -18.87 & 1.18 \\
706 & 0.174 & 0.023 & 53626.04 & 1.61 & 0.50 & 0.08 & 5.00 & 0.10 & 0.02 & 148.45 & 63.0 & -17.68 & 0.19 \\
714 &  &  &  &  &  &  &  &  &  &  &  &  &  \\
716 & 0.322 &  & 53639.38 & 0.13 & -0.04 & 0.01 & 5.00 & 0.00 & 0.03 & 3748.20 & 200.0 & -20.87 & 0.18 \\
717 & 0.129 &  & 53606.80 & 0.01 & 0.50 & 0.01 & -0.90 & 0.58 & 0.03 & 63.88 & 40.0 & -17.51 & 1.20 \\
722 & 0.085 &  & 53613.30 & 0.05 & -0.02 & 0.02 & -0.61 & 0.13 & 0.02 & 67.31 & 40.0 & -19.18 & 1.17 \\
735 & 0.189 &  & 53610.81 & 0.00 & -0.00 & 0.06 & -2.57 & 0.42 & 0.02 & 44.99 & 45.0 & -18.93 & 1.66 \\
739 & 0.106 &  & 53612.98 & 0.00 & 0.05 & 0.02 & -1.47 & 0.14 & 0.03 & 128.89 & 40.0 & -19.02 & 1.36 \\
744 & 0.127 &  & 53613.83 & 0.00 & 0.08 & 0.02 & 1.21 & 0.21 & 0.05 & 41.52 & 40.0 & -19.16 & 0.84 \\
746 & 0.179 & 0.187 & 53621.32 & 0.99 & -0.34 & 0.57 & 5.00 & 6.63 & 0.02 & 1265.86 & 62.0 & -18.64 & 0.18
\enddata
\end{deluxetable}
\end{longrotatetable}

\begin{deluxetable}{ccccc}
\tablecaption{Classification coordinates for objects classified as being photometrically similar to SN~1991bg. Included are coordinates calculated by fitting photometric bandpasses independently (xBand, yBand) and as collective red/blue sets (xCollective, yCollective). SNe are classified as being 91bg-like if they satisfy xCollective $x > 3$ and yCollective $ > 0$. Missing entries indicate a set of fits where one or more fits failed to converge.} \label{tab:classification_coords}
\tablehead{\colhead{CID} & \colhead{xBand} & \colhead{yBand} & \colhead{xCollective} & \colhead{yCollective}}
\startdata
2778 & 8.42 & 2.96 & 6.46 & 9.62 \\
11570 & -4.53 & -33.21 & 4.34 & 13.74 \\
12689 &  &  & 6.32 & 0.75 \\
15204 & -19.94 & 11.02 & 33.66 & 137.53 \\
16215 & -16.9 & -23.6 & 7.14 & 29.66 \\
16309 & -0.11 & -2.39 & 5.18 & 2.54 \\
16692 & 0.88 & -4.04 & 9.06 & 21.12 \\
17094 & 1.34 & 15.82 & 3.25 & 13.37 \\
17468 & 3.59 & 8.04 & 6.85 & 2.43 \\
17886 & 0.03 & 6.92 & 10.64 & 13.55 \\
18218 & 3.69 & 0.14 & 4.18 & 0.2 \\
18751 & 12.28 & 3.04 & 12.28 & 4.59 \\
18890 & 0.56 & 5.75 & 3.64 & 10.95 \\
19065 & -0.22 & 0.68 & 4.58 & 0.86 \\
21678 & 10.87 & 8.38 & 24.23 & 10.2 \\
21898 & 0.86 & 12.67 & 4.2 & 28.31
\enddata
\end{deluxetable}
\begin{deluxetable}{cccccc}
\tablecaption{A comparison of objects classified as anomalies by the machine learning classifier published in \citet{Pruzhinskaya19} and their corresponding classifications determined in this work. Objects with coordinates $x > 3.00, y > 0.00$ are classified as SN 1991bg-like SNe. Objects with coordinates $x < 3.00, y < 0.00$ are classified as normal SNe.} \label{tab:osc_comparison}
\tablehead{\colhead{CID} & \colhead{PSNID} & \colhead{Datasets} & \colhead{x} & \colhead{y} & \colhead{Class}}
\startdata
1706 & pSNII & 8 & -9.43 & -4.02 & Normal \\
2050 & Unknown & 7 & -0.72 & -0.26 & Normal \\
2093 & pSNII & 2 & -1.0 & -1.91 & Normal \\
2661 & SNII & 4 & 0.35 & -0.52 & Normal \\
2809 & pSNII & 2 & -1.43 & -0.52 & Normal \\
4226 & pSNII & 2 & -12.5 & -0.84 & Normal \\
4330 & pSNII & 5 & -1.66 & -1.28 & Normal \\
4652 & pSNII & 4 & -4.52 & -2.0 & Normal \\
5314 & pSNII & 8 & -8.71 & 0.97 & Peculiar \\
6992 & pSNII & 1 & -1.64 & -0.38 & Normal \\
12868 & pSNII & 3 & -2.62 & -1.02 & Normal \\
13112 & pSNII & 8 & -1.31 & -2.55 & Normal \\
13291 & pSNII & 4 & -47.91 & -20.57 & Normal \\
13461 & pSNII & 8 & -1.8 & -1.77 & Normal \\
13589 & pSNII & 4 & -2.62 & -0.63 & Normal \\
13725 & pSNII & 6 & -11.87 & -2.37 & Normal \\
13741 & pSNII & 5 & -1.13 & -0.64 & Normal \\
14170 & pSNII & 8 & -2.09 & -0.75 & Normal \\
15048 & pSNIa & 3 & -0.68 & -0.04 & Normal \\
15565 & pSNII & 8 & 0.97 & 0.16 & Peculiar \\
15745 & pSNIa & 3 & -0.56 & -0.05 & Normal \\
16302 & pSNIa & 3 & -4.37 & -3.21 & Normal \\
17292 & pSNII & 5 & -13.92 & -7.07 & Normal \\
17317 & zSNII & 2 & -2.91 & -5.5 & Normal \\
17339 & pSNII & 7 & -6.45 & -2.0 & Normal \\
17509 & pSNII & 6 & -23.32 & -4.59 & Normal \\
17756 & pSNII & 8 & -16.31 & -2.53 & Normal \\
17789 & SLSN & 2 & -40.1 & -27.26 & Normal \\
18228 & pSNII & 1 & -0.27 & 0.04 & Peculiar \\
18266 & pSNII & 3 & -31.69 & -6.44 & Normal \\
18391 & Unknown & 2 & 2.16 & -3.79 & Normal \\
18733 & pSNII & 1 & -2.33 & -2.06 & Normal \\
19047 & zSNII & 1 & -1.76 & -0.16 & Normal \\
19395 & pSNII & 1 & -2.36 & -0.28 & Normal \\
19504 & SNII & 3 & 0.18 & -0.03 & Normal \\
19699 & pSNII & 3 & -2.26 & -0.66 & Normal \\
20266 & pSNII & 4 & -1.62 & -0.81 & Normal
\enddata
\end{deluxetable}

\end{document}